\documentstyle[11pt,aaspp,flushrt,psfig,epsf]{article}

\begin{document}

\def\beq{\begin{equation}}
\def\eeq{\end{equation}}
\def\eq{equation \ }
\def\eqs{equations \ }


\title {THE EVOLUTION AND LUMINOSITY FUNCTION OF QUASARS FROM
	COMPLETE OPTICAL SURVEYS}
\author{Alexander Maloney\altaffilmark{1,2} and
Vah\'{e} Petrosian\altaffilmark{3,4}}

Center for Space Science and Astrophysics, Stanford University,
Stanford, CA 94305

\authoraddr{Varian 302c, mail code 4060, Stanford University,
Stanford, CA 94305}


\altaffiltext{1}{Departments of Physics and Mathematics}
\altaffiltext{2}{email: maloney@bigbang.stanford.edu}
\altaffiltext{3}{Departments of Physics and Applied Physics}
\altaffiltext{4}{email: vahe@bigbang.stanford.edu}

\begin{abstract}
We use several quasar samples (LBQS, HBQS, Durham/AAT and EQS) to determine 
the density and luminosity evolution of quasars.  Combining these different 
samples and accounting for varying selection criteria require tests of 
correlation and the determination of density functions for multiply truncated 
data. We describe new non-parametric techniques for accomplishing these tasks, 
which have been developed recently by Efron and Petrosian (1998). With these 
methods, the luminosity evolution can be found through an investigation of the
correlation of the bivariate distribution of luminosities and redshifts. We 
use matter dominated cosmological models with either zero cosmological 
constant or zero spatial curvature to determine luminosities from fluxes. Of 
the two most commonly used models for luminosity evolution, 
$L \propto e^{k t(z)} $ and $L \propto (1+z)^{k'}$, we find that the second 
form is a better description of the data at all luminosities; we find 
$k' =  2.58 $ ($[2.14,2.91]$ one $\sigma$ region) for the Einstein - de Sitter
cosmological model.

Using this form of luminosity evolution we determine a global luminosity 
function and the evolution of the co-moving density for the two types of 
cosmological models. For the Einstein - de Sitter cosmological model we find a
relatively strong increase in co-moving density up to $z \lesssim 2$, at which
point the density peaks and begins to decrease rapidly.  This is in agreement 
with results from high redshift surveys. We find some co-moving density 
evolution for all cosmological models, with the rate of evolution lower
for models with lower matter density. We find that the local cumulative 
luminosity function $\Phi(L_o)$ exhibits the usual double power law behavior.
The luminosity density ${\cal L} (z) = \int_0^\infty L \Psi(L,z) dL$, where 
$\Psi(L,z)$ is the luminosity function, is found to increase rapidly at low 
redshift and to reach a peak at around $z \approx 2$. Our results for 
${\cal L} (z)$ are compared to results from high redshift surveys and to 
the variation of the star formation rate with redshift.
 
\end{abstract}

\keywords{Quasars, Luminosity Function, Cosmology}

\section{INTRODUCTION}

Investigations of the evolution of the quasar population have
played a major role in the development of our ideas about the nature of 
these sources and their connection to other extragalactic objects.
Ever since the first complete survey of 3C radio quasars by
Schmidt (1968) and the subsequent survey of 4C quasars by Lynds and
Wills (1972) it has been evident that the population of quasars as 
a whole has undergone rapid evolution.  Using the so-called
$\langle V / V_{max} \rangle$ method these authors interpreted the evolution
with redshift $z$
as caused by an increase in the co-moving density of quasars with redshift.
However, 
both the source counts (Giaconni et al. 1979, Tananbaum et al. 1979)
and the redshift distribution of optically selected samples of quasars
(see e.g. Marshall 1985) clearly showed that such pure density evolution
(PDE) models, for which the luminosity function is separable as
\beq\label{eq:pde}
\Psi(L,z) = \psi(L) \rho (z) ,
\eeq
cannot be correct.  As more data was accumulated the pure luminosity 
evolution (PLE) model, with
\beq
\Psi(L,z) = \psi(L/g(z)) / g(z) ,
\eeq
gained more popularity.  The function $g(z)$ describes the luminosity
evolution of the population and $L_o = L/g(z)$ is the luminosity adjusted
to its present epoch ($z=0$) value.  
This model, while providing a better fit to the data than that of pure
density evolution, also appears to be inadequate in many cases
(see e.g. Petrosian 1973, Schmidt and Green 1978, Koo and Kron 1988,
Caditz and Petrosian 1990). 
Without loss of generality, we can write the luminosity function
as 
\beq\label{eq:gle}
\Psi(L,z) = \rho (z) \psi(L/g(z),\alpha_i) / g(z) .
\eeq
With $\psi$ normalized such that $\int_0^\infty \psi(L,\alpha_i) dL=1$,
$\rho (z)$ gives the density and its evolution and $\psi (L_o, \alpha_i)$
describes the local luminosity function (with $g(0) = 1$).  Here we explicitly
include the $\alpha_i$, parameters such as spectral index and break luminosity
that describe the shape of the luminosity function.  In general,
these parameters may vary with redshift.  A surprising, and a priori 
unexpected, result has been the absence of evidence for strong shape 
variation.  

Such results imply that the density and luminosity functions
$\rho (z)$ and $g(z)$ describe the {\it physical evolution}
of sources; e.g. the rate of birth and death of 
sources and the
changes in source luminosity with time.  Cavaliere and colleagues (see
Cavaliere and Padovani 1988 and references cited therein) were the first
to emphasize this fact.  A more complete description of the relation 
between the physical evolution and the functions describing the
generalized luminosity function (or statistical evolution) can be found
in Caditz and Petrosian (1990).  Unfortunately, this relation is not unique,
thus any test of quasar models via their expected evolution with cosmic
epoch must usually involve additional assumptions.  For example,
quasars could be long lived (compared to the Hubble time) sources created
during a relatively short period at high redshift undergoing continuous
luminosity evolution.  This is the model used in Caditz, Petrosian and Wandel 
(1991).  Alternatively, quasars could be short lived phenomena with
birth rate, death rate and luminosity that vary systematically with redshift
(see Siemiginowska and Elvis 1997).
Most of the work along these lines assumes what is now the standard
model for the source of energy of quasars and other active galactic 
nuclei (or AGNs); namely, accretion onto a massive black hole.  

Another interesting aspect of quasar evolution is the relation between the
evolution of the luminosity density of quasars, 
${\cal L} = \int L \Psi(L,z) dL$, and similar 
functions describing the evolution 
of galaxies, such as the star formation rate (SFR).
As shown by the high
redshift surveys of Schmidt et al. (1995) and others 
(e.g. Warren et al. 1994), 
the rapid rise with redshift of
the luminosity of quasars stops around redshift $2$ or $3$ and is expected to
drop, perhaps mimicking the SFR (see Shaver et al. 1998, 
Cavaliere and Vitorini 1998, and Hawkins and V\'{e}ron 1996).

In this paper we determine the luminosity function and its evolution for 
combined samples of quasars from various surveys, described in 
$\S 2$.  For a complete review of the various ways of accomplishing this task,
see Petrosian (1992).  We use here non-parametric methods based on
Lynden-Bell's (1971) idea which was generalized by Efron and Petrosian
(1992) to allow not only a determination of the functional forms but also 
a test of
correlation between redshifts and luminosities. This is essential for any
determination of the functional form of the luminosity evolution $g(z)$.
The above papers deal with magnitude limited samples with only an upper 
magnitude (lower flux) limit.  However, because of observational 
constraints, some of the samples are truncated in redshift space
and some have an upper magnitude (lower flux) limit.
The methods described in the above papers cannot be used for such multiply
truncated data.  New methods for treating this type of data 
were developed by Efron and Petrosian (1998).  In $\S 3$
we describe these new techniques and their relation to the older 
methods.  The choice of cosmological model also plays a crucial role in
such determinations.  It is well known that one can not determine both the
evolution of the luminosity function and the parameters of the 
cosmological model from magnitude limited samples alone.  Only by assuming
values for the cosmological parameters such as matter density, curvature and
cosmological constant can one calculate the form of $\Psi(L,z)$.
Alternatively, with some assumptions about $\Psi(L,z)$ one can test 
various cosmological models.  In $\S 4$ we describe the cosmological model 
parameters and in $\S 5$ we present the results of applications of the 
new statistical methods to the data described in $\S 2$.
Finally, in $\S 6$ we summarize these results and present our conclusions.

\section{THE DATA}\label{sec:data}

There have been many surveys of quasars, ranging from the original
radio selected samples of the 3C (Schmidt 1963) and 4C 
(Lynds and Wills 1972) surveys to a variety of other optically and
X-ray selected samples.  In this paper, we will focus only
on optically selected
data.  We use data from four samples with relatively similar 
selection criteria 
that provide a somewhat homogeneous data set spanning a large area of the
$L-z$ plane.  In order to combine the samples, all 
magnitudes were transformed to the $B$ band.
Some of the samples were artificially truncated in order to insure 
completeness within the truncation limits.
The combined sample consisted of 1552 objects in 
the range $ 15.5 \lesssim B \lesssim 21.2 $ 
and $0.3 < z < 2.2$, with well defined upper
and lower magnitude limits for each object. 

Figure $1$ shows the distribution of the complete surveys in the $B-z$ plane,
which at first sight shows little evidence for a Hubble relation of
$B \propto 5 \log (z)$.  However, as we shall discuss below, there 
is evidence for cosmological dimming with redshift of the sources.

\begin{figure}[htbp]
\leavevmode\centering
\psfig{file=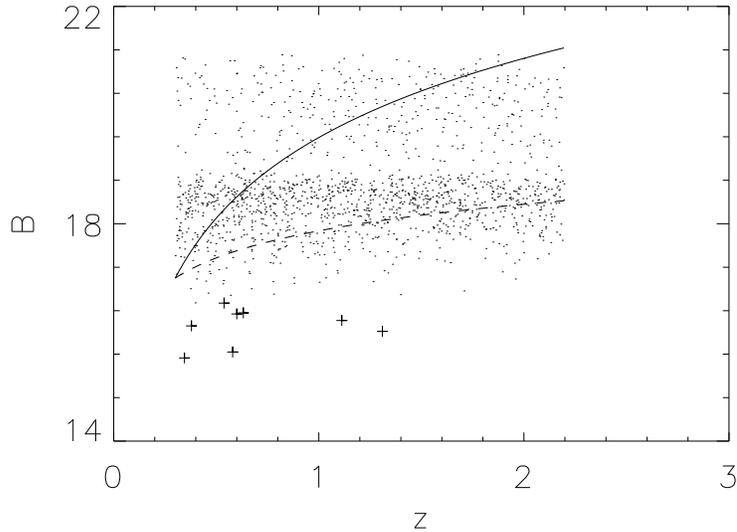,width=0.7\textwidth,height=0.5\textwidth}
\caption{The B magnitude - redshift data for the four samples LBQS, HBQS,
Durham/AAT and EQS described in $\S 2$, for $0.3 < z < 2.2$.
The EQS data are marked with plus signs.  The data are fitted to the 
parametric form $B(z) = B - \beta \log ( d^2_L(z) K(z)) + $ constant,
where $d_L(z)$ is the luminosity distance and $K(z)$ is the K-correction
term. 
The best fit ($\beta = 0.84$ for the Einstein - de Sitter cosmological
model) is shown by the dashed line.
The solid line shows
the expected relation between $B$ and $z$ for standard candle 
sources without luminosity evolution ($\beta = 2.5$) in the Einstein -
de Sitter model.
}
\end{figure}

\subsection{The Large Bright QSO Survey}
The Large Bright QSO survey (LBQS) contains 1055 QSOs in the
magnitude range $16.0 < B_J < 18.85$ and redshift range $0.2 < z < 3.4$
(Hewett et al. 1995).
The faint magnitude limits for the 18 fields range from $18.41$ to $18.85$
and the bright magnitude limit is $16.0$ for the entire sample.  In order to
transform from $B_J$ to $B$ magnitudes the color equation of Blair and Gilmore 
(1982)  
\beq\label{bj}
B = B_J + 0.28 (B - V)
\eeq
was used assuming an average $B-V$ of $0.3$ for the entire sample.  
When we combine this sample with the AAT sample and others we introduce
artificial cutoffs at $z = 0.3$ and $z = 2.2$,
which are the completeness limits for the AAT data.  
In addition, we removed all objects brighter than $B = 16.5$ 
in order to insure completeness near the bright magnitude limit,
leaving a total of 871 objects.

\subsection{The Homogeneous Bright QSO Survey}
Data from the six deepest fields of the Homogeneous Bright QSO Survey (HBQS)
were published by Cristiani et al. (1995) 
The six deepest fields of the HBQS contain 
either $B_J$ or $B'$ magnitudes for 285 QSOs in the
range $ 15.5 < B_J < 18.85$, with faint
magnitude limits ranging from $18.25$ to $18.85$.  The bright magnitude limits
are generally lower than those of the LBQS, varying from $14.0$ to $16.0$.
For observations in the $B_J$ band equation (\ref{bj}) was
used and for observations in the $B'$ band the equation
\beq
B = B' + 0.11 (B-V)
\eeq
of Blair and Gilmore (1982) was used assuming $B - V = 0.3$.  
Again, when combining this sample only objects in the redshift range 
$0.3 < z < 2.2$ were used, leaving a total of $254$ objects.

\subsection{The Durham/AAT Survey}
The Durham/AAT survey contains $419$ QSOs in the magnitude range 
$17 < b < 21.27$ (Boyle et al.	 1990) and redshift range
$0.3 < z < 2.2$, giving information about the QSO
luminosity function in a different regime than the previous two samples.
The faint magnitude limit ranges
from $20.25$ to $21.27$ and the bright magnitude limit ranges from
$16.4$ to $18.0$.  The data are given in the ``$b$'' magnitude system,
which according to Boyle et al. (1990) may be converted into the 
$B$ system by the relation
\beq
B = b + 0.23 (B - V - 0.9) .
\eeq
As above, the average value $B-V = 0.3$ was used to determine $B$.  

\subsection{The Edinburgh QSO Survey}
A subsample of the Edinburgh QSO Survey (EQS) consisting of $12$ QSOs
brighter than $B = 16.5$ was published by Goldschmidt et al. in 1992.
Of these, the $8$ that fall in the redshift range $0.3 < z < 2.2$
were added to the combined samples to give information about the 
luminosity function at the bright end. 

There have been several previous analyses of the above quasar
surveys.  Boyle et al. (1990) used binning techniques to fit the
the Durham/AAT data to the PLE model.  More recently, 
La Franca and Cristiani (1996) fit the LBQS and HBQS data to a 
more complex luminosity function (involving 3 or more parameters)
with no density evolution.  Hatziminaoglou et al. (1998) examined the
PLE and PDE cases separately using both the Durham/AAT and LBQS data.
As described below, we combine all of these data and determine
both the luminosity evolution $g(z)$ and the density evolution 
$\rho(z)$.

\section{STATISTICAL METHODS}\label{sec:statistics}
The statistical problem at hand is the determination of the true distribution
of luminosities and redshifts of the sources from a biased or truncated
data set, such as a flux limited sample.  Several different techniques 
exist that give such a determination.
The most common method is to bin the data and fit to some
parametric forms of $\psi(L)$ or $\rho(z)$.  However, it is preferable to avoid
binning and to use non-parametric methods whenever possible.  For a review 
of the various methods see Petrosian (1992).  Assuming the general form 
of equation (\ref{eq:gle}) for the luminosity function we must determine the
functional forms of the local 
luminosity function $\psi(L_o,\alpha_i)$, density evolution
$\rho(z)$ and the luminosity evolution $g(z)$ as well as the
changes in the parameters $\alpha_i$ with redshift, if any.  
This last aspect is a higher
order effect and will not be within the scope of this paper.  We will
discuss the certainty with which this can be ignored in the final analysis.

All non-parametric techniques for determining the distribution in a 
bivariate setting
require that the data be expressed in terms of two uncorrelated 
variables, i.e. that we use variables $x$ and $y$ for which
the density function is separable:
$\Psi (x,y) = \rho(x) \psi(y)$.  Thus before applying  non-parametric 
methods one must first determine the degree of correlation 
of the data in the $x - y$ plane.
This determination and the process of removing the correlation is
equivalent to a determination of the functional form of the luminosity
evolution $g(z)$ and the subsequent transformation $L \to L_o = L / g(z)$.
This cannot be accomplished easily by non-parametric methods.  Therefore,
we chose two parametric forms for the luminosity evolution, 
$g_k(z)$ and $g_{k'}(z)$, and find the values of the parameters 
$k$ and $k'$ for which $L_o$ and $z$ are uncorrelated.
Once this is done, the non-parametric methods described in Petrosian (1992)
may be used to determine $\rho(z)$ and $\psi(L_o)$.

The methods normally used to test correlation and determine
the distributions are suited for simple truncations, such as
$y < f(x)$ (which by defining $x' = f(x)$ can be reduced to 
the generic case of $y < x'$).  
This is sufficient for simple flux limited data.  However,
the majority of astronomical data, and quasar data in particular,
suffer from more than one truncation.  The data may have an upper
as well as a lower truncation, $f^-(x) < y < f^+(x)$, or there may be similar
truncations in the value of the other variable.  In addition, the functions
$f^-(x)$ and $f^+(x)$ may not be continuous or even single valued. 
In general, multiply truncated redshift - magnitude data may be written 
as $\{z_i,m_i, [z^-_i,z^+_i], [m^-_i,m^+_i]\}_{i=1}^N$ where $[z^-_i,z^+_i]$
and $[m^-_i,m^+_i]$ are the observational
limits on $z$ and $m$ for the $i^{th}$ object,
respectively.  Given a cosmological model $\Omega$ this gives data of the form
$\{z_i,L_i, [z^-_i,z^+_i], [L^-_i,L^+_i]\}_{i=1}^N$.
If one assumes that each object has the same redshift limits $[z^-_i,z^+_i]$,
which is the case in the majority of our analyses,
then the problem is to test the correlation and distribution of
$x$ and $y$ from a data set $\{x_i,y_i\}_{i=1}^N$ given truncation limits
$[y^-_i,y^+_i]$ for each point.
The previous methods developed for this test (see Petrosian 1992 and Efron
and Petrosian 1992) are suited for one sided truncations.  In a more
recent work (Efron and Petrosian 1998) we have developed methods, which are
a generalization of the earlier methods, for dealing with doubly truncated
data.  We will briefly 
review these new methods of testing correlation and non-parametrically 
determining the density evolution and luminosity function.

\subsection{Tests of Correlation and Determination of Luminosity Evolution}

\subsubsection{Untruncated data}
If $x$ and $y$ are independent then the rank $R_i$ of $x_i$ in an 
untruncated sample (i.e. a sample truncated parallel to the $x$ and $y$ 
axes, so that $y_i^\pm$ are independent of $x_i$) 
will be distributed uniformly 
between $1$ and $N$ with an expected mean $E={1 \over 2} (N+1) $ and variance
$V = {1 \over 12} (N^2 - 1)$. 
One may then normalize $R_i$ to have a mean of $0$ and a variance of $1$
by defining the statistic $T_i =  (R_i - E) / V $.
One then rejects or accepts the hypothesis of independence based on the
distribution of the $T_i$.

One simple way of doing so is by defining a 
single statistic $t_{data}$ based on the $T_i$ with a mean of $0$ and 
a variance of $1$.
One then rejects the hypothesis of independence if $|t_{data}|$ is too large 
(e.g. $|t_{data}| \geq 1$ for rejection of independence at the
$ 1 \ \sigma$ level).  The quantity
\beq\label{taudef_1a}
\tau = { {\sum_i ( R_i - E )} \over \sqrt{\sum_i V} }
\eeq
is one choice of such a test statistic.  This $\tau$ is equivalent to
Kendell's $\tau$ statistic, which is defined in the following manner.  
Consider all possible pairings 
${\cal P} = \{(i,j)\}$ between data points and call
a pairing $(i,j)$ positive if $(x_i - x_j) (y_i - y_j) > 0 $ and
negative if $(x_i - x_j) (y_i - y_j) < 0 $.
If there are no ties then the $\tau$ of equation (\ref{taudef_1a})
is equivalent to Kendell's $\tau$ statistic
\beq\label{eq:taudef_2}
\tau = {  \# {\rm \ positive \ } (i,j) \in {\cal P} - 
\# {\rm \ negative \ } (i,j) \in {\cal P}
	\over \# \ (i,j) \in {\cal P} }.
\eeq

\subsubsection{Data with One-Sided Truncation}
A straightforward application of this method to truncated data will clearly
give a false correlation signal.   Efron and Petrosian (1992), and 
independently Tsai (1990), describe how this method can be applied to
data with one-sided truncation, i.e. either $y_i^-= -\infty,\ i=1,\ldots, N$ or
$y_i^+ = \infty, \ i=1,\ldots,N$.  For example, if $y_i^+ = \infty$ but 
$y_i^-$ varies with $x_i$ then the above procedure is modified as follows.
For each object define the {\it comparable} or {\it associated} set
\beq\label{eq:jdef}
J_i = \{ j : y_j > y_i, \ y_j^- < y_i \}
\eeq
to consist of all objects of greater $y$ for which the value $y=y_i$ could
possibly be observed.  This is the same set as defined in Lynden-Bell's 
$C^-$ method and, unlike the definition of Efron and Petrosian 
(1992, e.q. (2.9)), does not include the object in question. 
In the case of luminosity-redshift data this
is the largest subset of luminosity (not magnitude or flux) and volume 
limited data that can be constructed for each given $(L_i, z_i)$.
If $x$ and $y$ are independent then we expect the rank $R_i$ of
$x_i$ in the eligible set, 
\beq\label{Ri}
R_i = \# \{j \in J_i : x_j \leq x_i\},
\eeq
to be uniformly distributed between $1$ and $N_i$, where
$N_i$ is the number of points in $J_i$.
The rest of the procedure follows as for untruncated data.
The normalized statistic $T_i$ is defined here as 
$T_i = (R_i - E_i) / V_i$ where $E_i = {1 \over 2} (N_i + 1)$ and
$V_i = {1 \over 12} (N_i^2 - 1)$.  The test statistic $\tau$
is then defined by
\beq\label{taudef_1}
\tau = { {\sum_i ( R_i - E_i )} \over \sqrt{\sum_i V_i} }
\eeq
and is equivalent to a
version of Kendell's $\tau$ statistic defined by equation
(\ref{eq:taudef_2}).  In this case, however, we consider only the
set of possible pairings allowed by truncation
${\cal P} = \{(i,j): y_i > y_j^-, y_j > y_i^-\}$. 

\subsubsection{Multiply Truncated Data}
A generalization of the above method to doubly (or multiply) truncated data
was developed recently by Efron and Petrosian (1998).  The method is 
equivalent to the previous method, with the eligible set defined as
\beq
J_i = \{ j : y_j > y_i, \ y_i \in (y_j^-, y_j^+)\}
\eeq
and the set of allowed pairings
\beq
{\cal P} = \{ (i,j) : y_i \in (y^-_j,y^+_j),y_j \in (y^-_i,y^+_i) \}
\eeq
defined such that each object lies within the truncation region of the other.

In this case, however, the distribution of the rankings (or of $\tau$)
is unknown.  If the data
are uncorrelated then $\tau$ must still have a mean of zero
and a bootstrap method may be used to determine the variance $V_\tau$ 
as follows.  By assuming the
data $\{x_i,y_i\}$ are uncorrelated one can use the methods of the
next section to determine $\psi(y)$, the underlying (i.e. non-truncated)
probability distribution of $y$.  
Once $\psi(y)$ is found one can simulate $N_{sim}$ sets of
data with underlying probability density
$\psi(y)$ and truncation limits $[y^-_i,y^+_i]$ for the $i^{th}$ object
in each set.  
For each simulated set of data ${\cal D}_k$ one may find $\tau_k$ as in 
equation (\ref{eq:taudef_2}) and estimate
$V_\tau$ from the distribution of $\{ \tau_k \}_{k=1}^{N_{sim}}$.  For large
numbers of simulations $N_{sim}$ the error in this determination
of $V_\tau$ is approximately $V_\tau / \sqrt{N_{sim}}$.  Given 
$V_\tau$ one can define a normalized test statistic 
$\tau / V_\tau$ with a mean of $0$ and a variance of $1$.

\subsubsection{The Luminosity Evolution}
If $x$ and $y$, in this case the luminosity and redshift, prove to be 
independent, which would be the case if 
$|\tau| < 1$, one may assume that there is no
luminosity evolution and proceed with the determination of
the univariate distributions
$\psi(L)$ and $\rho(z)$ of equation (\ref{eq:pde}) using the methods
described below.  However, if $|\tau| \geq 1$ then 
$L$ and $z$ cannot be considered independent
and one may assume that the most likely explanation
is the presence of luminosity evolution ($g(z) \neq $ constant).  Another
possibility is the variation of the shape parameters $\alpha_i$ with $z$.  We
will return to this possibility below.  One can determine the function
$g(z)$ parametrically as follows.

Given a parametric form for luminosity evolution $g_k(z)$ one can 
transform the luminosities into $L_o (k) = L / g_k(z)$ and proceed with the
determination of the correlation
$\tau(k)$ between $L_o$ and $z$ as a function of the parameter $k$.  
If $\tau$ is normalized to have a standard deviation of $1$, then
the values of $k$ allowed
by the data at the $1 \ \sigma$ confidence level are
$\{ k : |\tau(k)| < 1 \}$, and the most likely values of $k$
are those with $\tau(k) = 0$.  Although in principle it is possible for
the function $\tau (k)$ to have several zeroes, this did not 
occur for the specific cases described in $\S 5$.

\subsection{Non-Parametric Determination of Distribution Functions}
Once a function $g$ is found that removes the correlation between $x$
and $y$ (or $z$ and $L$), 
the task is to find the underlying distributions
$\rho(x)$ (the density evolution) and $\psi(y')$ 
(the local luminosity function) given uncorrelated data 
$\{ x_i,y'_i = y_i / g(x_i) \}_{i=1}^N$ and truncation limits 
$[y'^-_i,y'^+_i]$.
If the truncation is one-sided (e.g. $y'^+_i = \infty$ for all $i$)
then a variety of non-parametric methods can be used to determine the
univariate distribution functions. 
As shown by Petrosian (1992) all non-parametric methods 
reduce to Lynden-Bell's (1971) method.  As with the tests of correlation 
described above, the gist of this
method is to find for each point the comparable
set defined in equation (\ref{eq:jdef}) and the number $N_i$ of points in 
this set. For example, the cumulative distribution in $y'$, 
$\Phi(y') = \int^\infty _{y'} \psi(t) dt $, is given by
\beq
\ln \Phi(y'_i) = \sum _{j < i} \ln (1 + {1 \over N_j}) . 
\eeq
For doubly truncated data the comparable set is not completely observed, thus
a simple analytic method such as the one described above is not possible.
However, it turns out that a simple iterative procedure can lead to a 
maximum likelihood estimate of the distributions.  Here we give a brief 
description of this method; for details the reader is referred to Efron 
and Petrosian (1998).

Assume 
that the underlying density function $\psi(y)$ is discretely distributed 
over the $N$ observed values of $y$.  If we let $\psi_i = \psi(y_i)$ 
be the probability density at $y_i$ then $\Phi_i = \sum_j \psi_j$, where the
summation includes all data points for which $y_j \in [y_i^-, y_i^+]$, 
is the total probability density for the 
truncation region $[y^-_i,y^+_i]$.  
If we define the matrix $\bf J$ by 
\beq\label{Jdef}
J_{ij} = \cases{1,&if $y_j \in [y^-_i,y^+_i] $\cr
		0,&if $y_j \notin [y^-_i,y^+_i] $\cr}
\eeq
then the definition of $\Phi_i$ is equivalent to 
${\bf \Phi} = {\bf J} \cdot {\bf \Psi}$ where 
${\bf \Psi} = (\psi_1,\dots,\psi_N)$ and
${\bf \Phi} = (\Phi_1,\dots,\Phi_N)$.  
This matrix ${\bf J}$ contains all of the information about the 
data and the truncation limits needed to find the 
vector of probability densities ${\bf \Psi}$.
We also have the normalization condition on ${\bf \Psi}$,
\beq\label{norm}
\sum_{i=1}^N \psi_i = 1 .
\eeq
From the definition of $\bf \Phi$ it follows that the conditional probability
$\psi(y_i|[y^-_j,y^+_j])$ of observing a value $y_i$ 
within the truncation region $[y^-_j,y^+_j]$ is 
\beq\label{condf}
\psi(y_i|[y^-_j,y^+_j]) =
   \cases{ \psi_i / \Phi_j,& if $y_i \in [y^-_j,y^+_j]$\cr
			0	,& if $y_i \notin [y^-_j,y^+_j]$\cr}.
\eeq
The final condition on ${\bf \Psi}$ is determined by maximizing the
likelihood of observing the actual data, 
\beq\label{prob}
P_{data} = \prod_{i=1}^N \psi(y_i|[y^-_i,y^+_i]) = 
\prod_{i=1}^N {\psi_i \over \sum _j J_{ij} \psi_j}.
\eeq
By setting
${\partial P_{data} \over \partial {\bf \Psi}} = 0$ it follows that
$\psi_k^{-1} = \sum_j J_{jk} \Phi_j^{-1}, k=1, \dots, N$.  This may be written 
compactly as
\beq\label{likely}
{1 \over {\bf \Psi}} = {\bf J}^\dagger {1 \over {\bf J} \cdot {\bf \Psi}}
\eeq
with the notation ${1 \over {\bf a}} = (a_1^{-1},\dots,a_N^{-1})$.
Thus we have reduced the problem of finding the density function for 
data with arbitrary truncation to the ``moral equivalent'' of an 
eigenvalue problem for the matrix $\bf J$.

In practice, this condition may be used as a recursive formula to determine
${\bf \Psi}$.  One starts with 
an initial guess for the density vector ${\bf \Psi}^o$.
Equation (\ref{likely}) then gives the recursion relation
\beq\label{recursive}
{1 \over {\bf \Psi}^{(j+1)}} = {\bf J^\dagger} 
{1 \over {\bf \Phi}^{(j)}} + c^{(j)}
\eeq
where the constant $c^{(j)}$ is determined by the normalization condition
$\sum_i \psi^{(j+1)}_i = 1$.  
One may use as an initial guess the untruncated solution 
$\psi_i^o = {1 \over N}$.  In most problems, however, one of the two 
truncations will have a more pronounced effect than the other.
In this case, one may ignore the weaker truncation and use the result based
on the one-sided method as an an initial guess.  
We found that with this initial guess and data
confined to one region of the $L - z$ plane the sequence of
${\bf \Psi}^{(j)}$ defined by equation (\ref{recursive}) usually converged 
quickly.  However, for combined samples spanning different regions of 
the $L - z$ plane it was helpful to use an algorithm to
accelerate the convergence of the series of ${\bf \Psi}^{(j)}$.
For this purpose we used Aitken's $\delta^2$ method, which 
gives an improved estimate for the terms of series by assuming 
approximately geometric convergence (see, e.g. Press et al. 1992, 
chapter 5.1).

\section{COSMOLOGY AND MODELS OF LUMINOSITY EVOLUTION}\label{sec:cosmology}

In order to determine the intrinsic parameters of an object from
the observed data one must assume a certain cosmological model.  
Many cosmological models may be described 
in terms of a few fundamental parameters, which (see e.g. Peebles 1993)
for a matter dominated
(non-relativistic) universe are the matter density $\rho_o$, 
the cosmological constant $\Lambda$ and the curvature of space 
$k$ (which is $+1$, $0$ or $-1$ for closed, flat or open universes,
respectively).
Using Hubble's constant $H_o$ these parameters may be written in
dimensionless form as
\beq
\Omega_M = {8 \pi G \rho \over 3 H_o ^2}, \,\, 
\Omega_k = - {k c^2  \over (H_o R_o)^2 } \,\,\,\, {\rm and} \,\,\,
\Omega_\Lambda = {\Lambda \over 3 H_o^2} 
\eeq
where $R_o$ is the value of the
expansion parameter of the universe at the present epoch.
These three parameters are related via the Friedman-Lemaitre equation
$\Omega_M + \Omega_k + \Omega_\Lambda = 1 $,
allowing us to eliminate one of them in favor of Hubble's constant.
For example, the curvature term may be written
\beq
\Omega_k = 1 - (\Omega_M + \Omega_\Lambda).
\eeq
We will consider the two classes of cosmological models given by
$\Omega_\Lambda=0$ 
(no cosmological constant) and 
$\Omega_k = 0$ 
(flat universe with cosmological constant).
For calculations of the luminosity function
we will pay particular attention to the two cases
$\Omega_k = \Omega_\Lambda = 0$ and $\Omega_\Lambda = 0.85$, $\Omega_M=0.15$.  
The first of these is the standard 
Einstein - de Sitter model, and the second is 
an inflationary model with parameters in accordance with current observations.
For definiteness, 
Hubble's constant was assumed to be $H_o = 70 \ {\rm km / (s \ Mpc)}$, although
most results are independent of this assumption.  

Once the values of the cosmological parameters are fixed, 
calculations of intrinsic parameters are 
relatively straightforward (Peebles 1993).
The absolute luminosity, for example, takes the form
\beq\label{eq:ldef}
L = f 4 \pi d_L^2 K (z)
\eeq
where $f$ is the observed flux, the luminosity distance $d_L$ is
\beq
d_L = {c \over H_o} (1 + z) 
{\sin {\sqrt k u(z)}
\over {\sqrt k} }
\eeq
and $K(z)$ is the K-correction term. 
The co-moving coordinate distance is
\beq
u(z) = \int_0^z
{dz \over \sqrt {\Omega_M (1+z)^3 + \Omega_k (1+z)^2 + \Omega_\Lambda}}
\eeq
and the co-moving volume contained within a sphere of radius 
corresponding to redshift $z$ is
\beq\label{eq:vdef}
V(z) = 4 \pi ({c \over H_o}) \int_0^z {du \over dz} {d_L^2 \over (1+z)^2} dz .
\eeq
In general, these integrals must be evaluated numerically.

In order to determine the K-correction term $K(z)$ one must
make an assumption about the quasar spectrum in the optical region.  
The general practice here is to
assume a power law spectrum $L_{optical} \propto \nu^\alpha$
with spectral index $\alpha \simeq -0.5$, which
gives $K(z) = (1+z)^{1+\alpha} \simeq \sqrt {(1+z)}$.

Two models for luminosity evolution were used: $g_k(z) \propto e^{kt(z)}$
and $g_k(z) \propto (1+z)^k$.  The first of these assumes an exponential
dependence on the fractional lookback time $t(z)$, which is defined as
$t(z) = 1 - {T(z) / T(0)}$, where $T(z)$ is the age of the universe 
at redshift $z$,
\beq
T(z) = H_o^{-1} \int_z^\infty 
{du \over dz} {dz \over (1+z)} 
.
\eeq
The second model assumes a power
law dependence on the scale factor of the universe (or the expansion parameter
$R$), which is independent of the cosmological parameters.  
Analyses of earlier data
(see e.g. Caditz and Petrosian 1990) have traditionally 
given estimates of quasar evolution
for these two parametric forms of $g(z) \approx e^{7.5 t(z)}$ and
$g(z) \approx (1+z)^{3}$.

\section{THE EVOLUTION OF THE LUMINOSITY FUNCTION}\label{sec:results1}

In what follows we apply the procedures of $\S 3$ to determine the 
correlation between the luminosities and redshifts and test parametric forms
for the evolution of the luminosity, the function $g(z)$.  Then
we transform all luminosities to their present epoch values
$L_o = L/g(z)$ and determine the co-moving density evolution $\rho(z)$
and the present epoch luminosity function $\psi(L_o)$
for the cosmological models described in $\S 4$.  We apply these tests
to the surveys described in $\S 2$ individually and in various combinations. 
Before presenting these results we discuss briefly the redshift - magnitude
data, i.e. the Hubble diagram shown in Figure $1$.

\subsection{The Quasar Hubble Diagram}

As is evident from Figure $1$, at first glance there seems to be very little
correlation between the redshifts and magnitudes (or fluxes) of quasars;
i.e. there is no obvious evidence for a Hubble type relation.  This result is
well known.  For example, a preliminary test of correlation between $m$ and
$z$ in a small subsample by Efron and Petrosian (1992, Figure 6) showed no
correlation.  Earlier, the absence of a clear Hubble relation was used as an
argument against the cosmological origin of quasar redshift 
(see, e.g. Burbidge and O'Dell 1973).  
This is not the only possible interpretation,
however.  One would not expect a simple Hubble diagram for sources with
a broad luminosity function (non-standard candle sources, see e.g. 
Petrosian 1974).  The absence of an obvious Hubble relation can also
arise from 
approximate cancelation between cosmological dimming and luminosity evolution. 
Exact cancelation of these two effects is highly implausible
and could bring into question the basic assumptions about the distribution
of sources.  To clarify this situation we have applied the correlation tests
described in $\S 3$ to the surveys of $\S 2$.  
First ignoring the high flux
(low magnitude) limit, i.e. with $m_{min} = - \infty$, we use the 
one-sided tests and find the results labeled $\tau_1$ shown in Table 1.
When using the double-sided tests we find the results $\tau_2$, which indicate 
slightly less correlation, as expected because the one-sided 
methods ignore the slight truncation induced by the high flux limits.
These tests, when applied to the combined data sets, give a
correlation of $\tau = 3.63$.  
This result rejects the hypothesis of independence between $B$ and $z$ at the
$ 99.97 \%$ confidence level and is independent of any
cosmological parameters.  In addition, we may test the parametric fit
$B(z) = B - \beta \log ( d^2_L(z, \Omega) K(z)) + $ constant for the data
using the methods for multiple truncations to determine the best value of
$\beta$, i.e. the value for which $B(z)$ and $z$ are uncorrelated 
($\tau = 0$).
A value of $\beta = 2.5$ is what one would expect for standard candle sources
with a very narrow luminosity function, while a value of $\beta = 0$ would 
mean the complete absence of a Hubble relation and the exact cancelation
described above.
The results shown in Table $1$ indicate that $\beta$, while clearly
less than $2.5$, differs significantly 
from $0$ for the cosmological models discussed in $\S 4$.  
The best parametric fit for the Einstein - de Sitter cosmological
model ($\beta = 0.84$) is shown in Figure 1 along with the expected
relation for standard candle sources ($\beta = 2.5$).

\begin{table*}
\begin{center}
\begin{tabular}{lrrrrrrr}
Sample & $N$ & $\tau_1$ & $P_1$ & $\tau_2$ & $P_2$ 
& $\beta_1$ & $\beta_2$ \\
\tableline
Durham/AAT & 419 & -0.65 & 48.71 & -0.45 & 35.02 & -0.22 & -0.18 \\
LBQS & 871 & 4.65 & 99.99 & 3.75 & 99.98 & 1.11 & 0.92 \\
HBQS & 254 & 1.37 & 82.92 & 1.27 & 79.42 & 0.71 & 0.59 \\
LBQS and HBQS & 1125 & 4.80 & 99.99 & 3.98 & 99.99 & 1.01 & 0.83 \\
Combined Data & 1552 & 3.98 & 99.99 & 3.63 & 99.97 & 0.84 & 0.70 \\
\end{tabular}
\end{center}

\tablenum{1}
\caption{Correlation data for the Quasar Hubble Diagram.  $N$ is the number 
of data points in each of the data sets of $\S 2$.  The correlation
$\tau$ and the probability value $P$ for rejection of independence 
between $B$ magnitude and redshift $z$ are 
given.  The first set of values ($\tau_1$, $P_1$) are found using the 
one-sided method of Efron and Petrosian (1992) and the second set of
values ($\tau_2$, $P_2$) are found using the general method for doubly
truncated data presented in $\S 3$.
The parameters $\beta_1$ and $\beta_2$ are found with this general method
by fitting to the equation $B(z) = B - \beta \log ( d^2_L(z) K(z)) \ + $ 
constant
for the two cosmological models $\Omega_M=1$, $\Omega_\Lambda=0$ and 
$\Omega_M=0.15$, $\Omega_\Lambda=0.85$ respectively.
}

\end{table*}

We now turn to a determination of the evolution of the luminosity.

\subsection{The Luminosity Evolution $g(z)$}
We examine 
the two commonly used parametric forms for the luminosity evolution, 
the exponential and power law forms.  These two different forms emphasize the
evolution in different regions of the luminosity $-$ redshift ($L-z$) plane.
A correct parameterization will have the same value for its parameters
when applied to samples with different limits (i.e. different coverage 
of the $L - z$ plane). This fact can be used to test a given parametric form.
Table 2 summarizes the results described in the subsequent sections. 
Figure 2 gives an example of
the variation of the test statistic $\tau$ as a function of 
the evolution parameter $k$.
The optimal value of $k$, 
i.e. the value which indicates that $L_o$ and $z$ are not
correlated, is given by the condition $\tau = 0$ 
and the $1 \ \sigma$ range of this parameter is given by the condition 
$|\tau| < 1$.  These values are shown in Figure 2.

\begin{table*}
\begin{center}
\begin{tabular}{lrrrrrrr}
Sample & $N$ & $k$ & $k_{min}$ & $k_{max}$ & $k'$ & $k'_{min}$ & $k'_{max}$ \\
\tableline
Durham/AAT & 419 & 8.72 & 6.66 & 10.07 & 3.53 & 2.57 & 5.05\\
HBQS & 254 & 5.39 & $- \infty$ & 6.49 & 3.20& $-\infty$ & 3.94\\
LBQS & 871 & 4.28 & 2.66 & 5.17 & 2.02 & 1.24 & 2.53 \\
Combined Data & 1552 &  5.15 & 4.36 & 5.70 & 2.58 & 2.14 & 2.91 \\
\end{tabular}
\end{center}

\tablenum{2}
\caption{ 
Values of the evolution parameters $k$ and $k'$ for the different data 
sets.  These parameters refer to luminosity evolutions
$g_k(z) \propto e^{kt(z)}$ and $g_{k'}(z) \propto (1 + z) ^ {k'}$,
respectively.  
The minimum and maximum values are those allowed at 
the $1 \ \sigma$ level.  Note the variation in $k$ and the near constancy
of $k'$.
}

\end{table*}

\begin{figure}[htbp]
\leavevmode\centering
\psfig{file=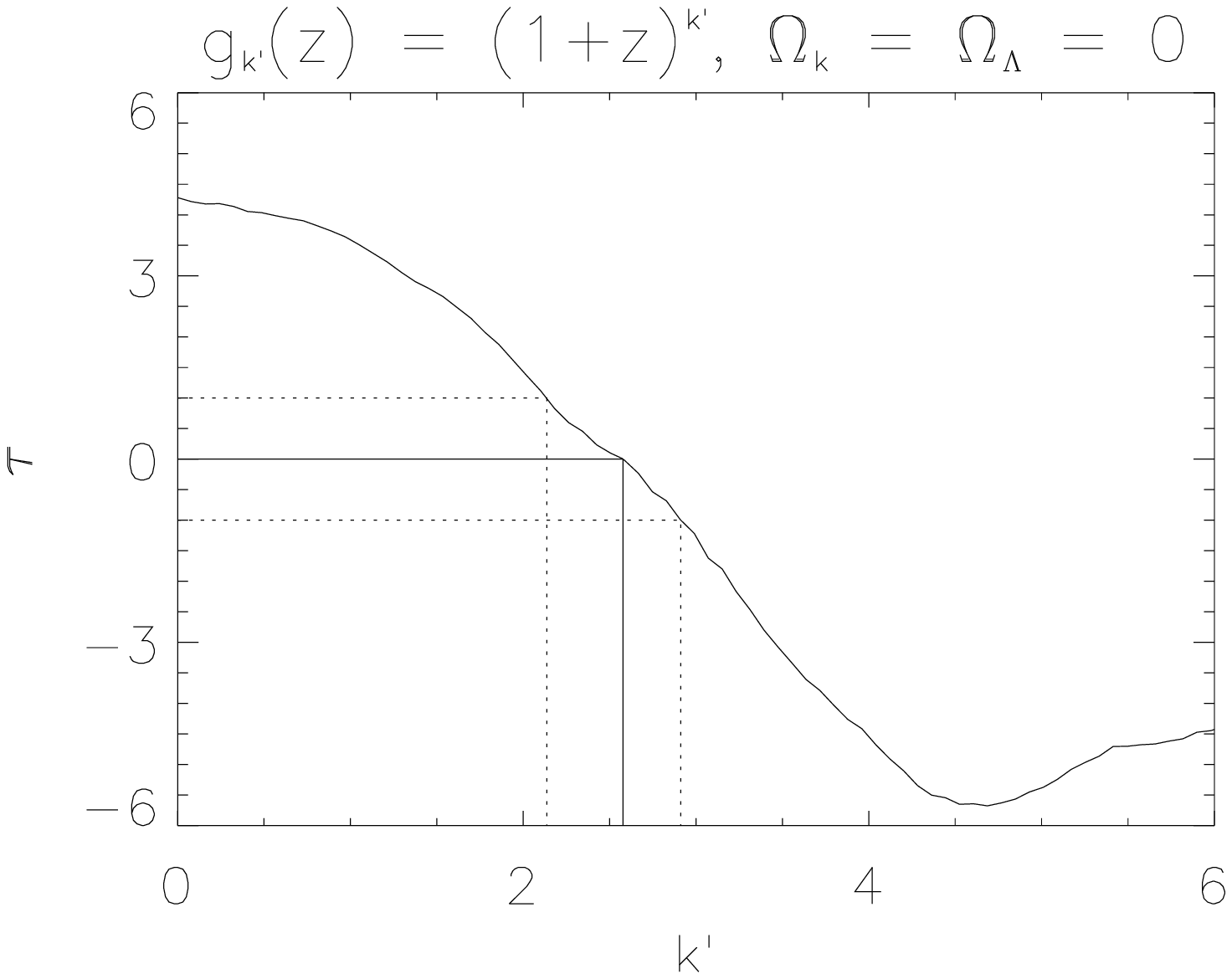,width=0.7\textwidth,height=0.5\textwidth}
\caption{A determination of the luminosity evolution parameter for the 
parametric form $g_{k'}(z) \propto (1 + z) ^ {k'}$. The correlation statistic 
$\tau$ is shown as a function of $k'$ for the combined data set for the 
Einstein - de Sitter cosmological model.  We use the correlation test for 
multiply truncated data of $\S 3$, normalized so that $\tau$ has a 
standard deviation of 1.  The solid line at $\tau = 0$ demonstrates 
the determination of the optimal value $k' = 2.58$ and the
dashed lines at $|\tau| = 1$ demonstrate the determination of the $ 1 \ \sigma$
region $[2.14,2.91]$.  
}
\end{figure}

\subsubsection{Evolution of the form $g_k(z) \propto e^{k t(z)}$}
Figures 3a and 3b display the best values and the $1 \ \sigma$ ranges
for the evolution parameter $k$ assuming both $\Omega_\Lambda = 0$ and 
$\Omega_k = 0$ cosmological models for the different samples.  
The Durham/AAT data exhibit much stronger evolution 
than the other data, indicating that this form of evolution 
does not adequately describe the data in the entire $L - z$ plane.
The values of $k$ found for most of the data sets are considerably less
than the previously determined value $k \approx 7.5$ by 
Caditz and Petrosian (1990), 
which was dominated by the Durham/AAT data.

\begin{figure}[htbp]
\leavevmode
\centerline{
\psfig{file=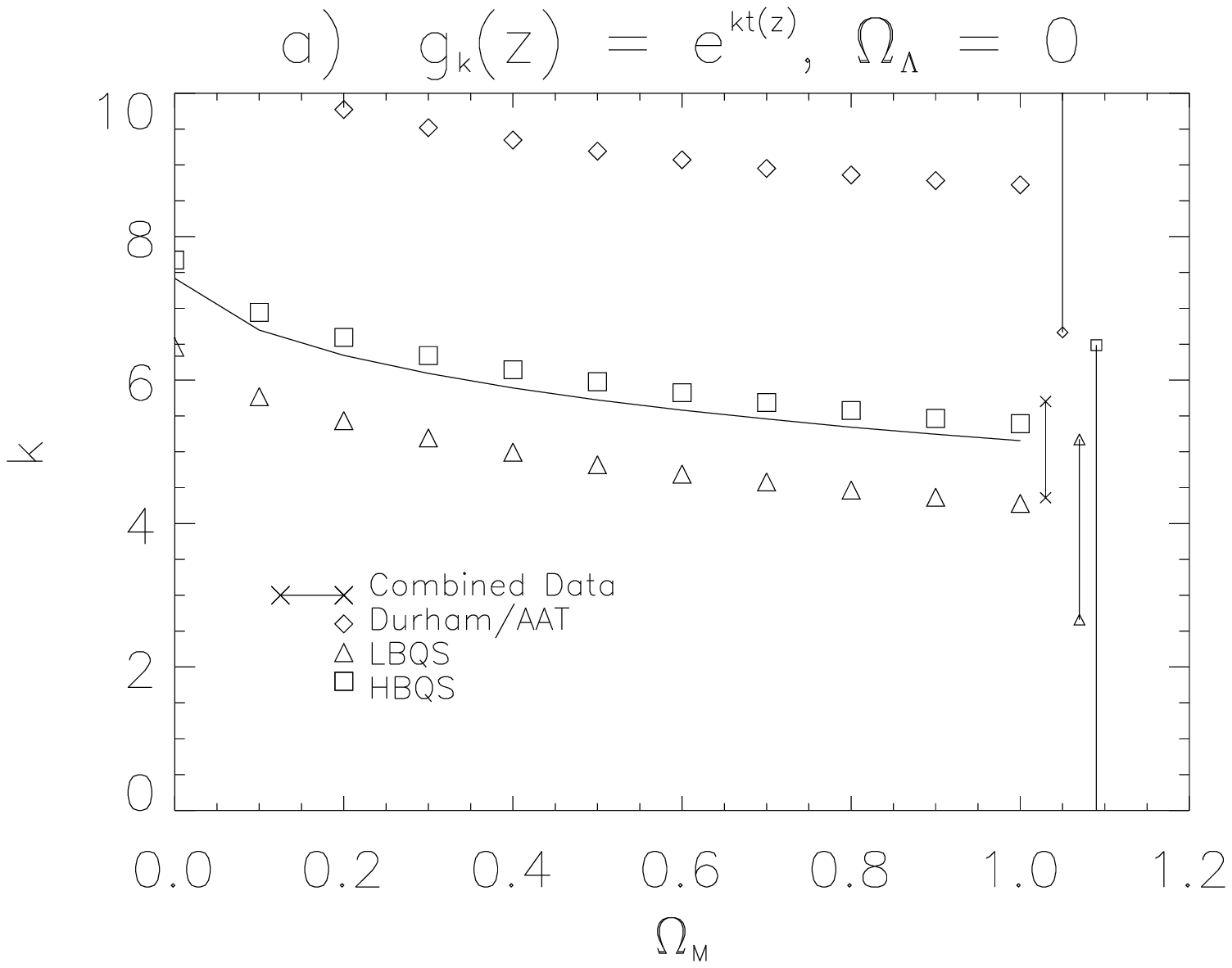,width=0.5\textwidth,height=0.4\textwidth}
\psfig{file=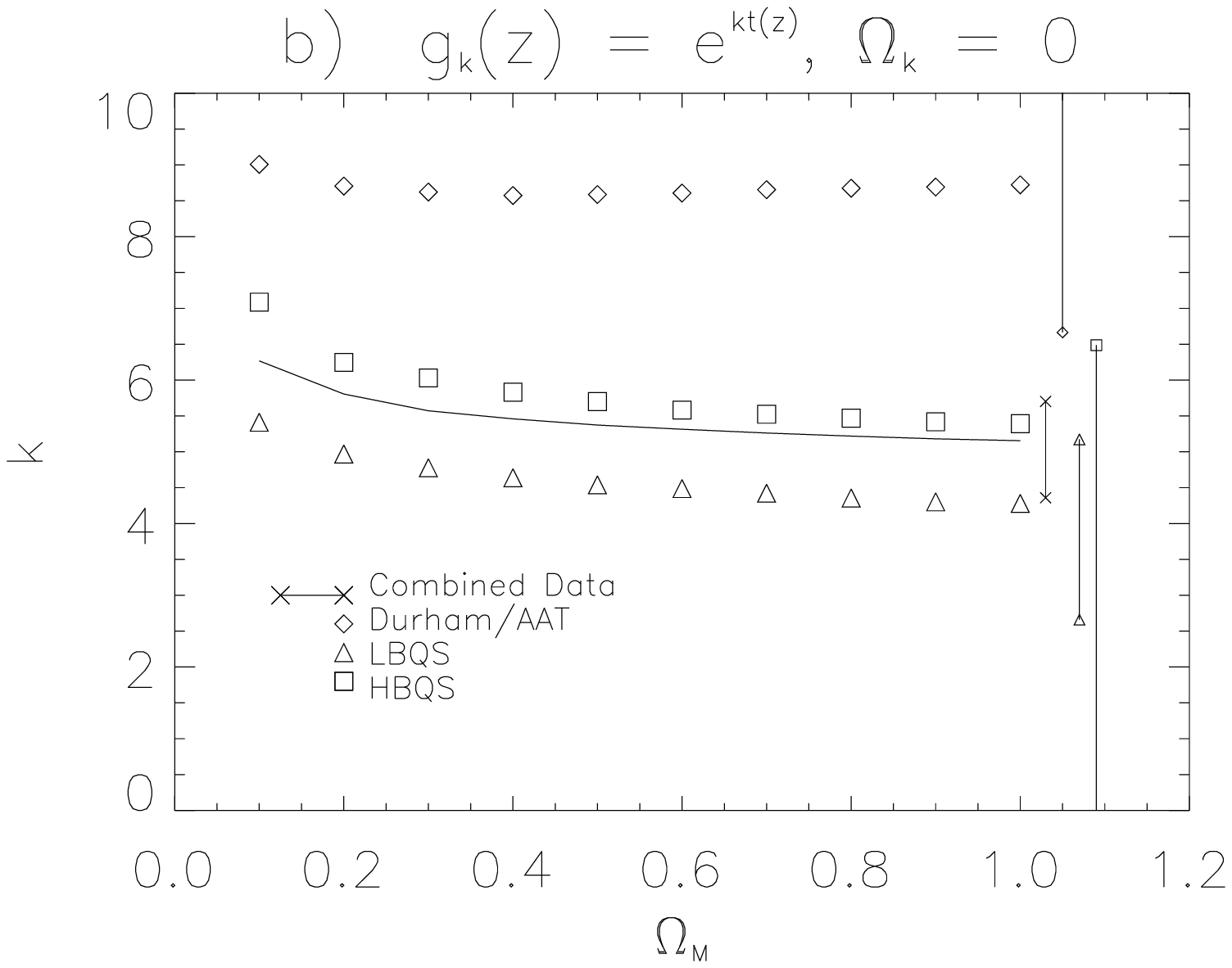,width=0.5\textwidth,height=0.4\textwidth}
}
\caption{Variation of the evolution parameter $k$ with $\Omega_M$ for
evolution of the form $g_k(z) \propto e^{kt(z)}$ for the three major surveys
and for the combined data set.  One $\sigma$ error bars are plotted for the 
$\Omega_M = 1$ case.  Error bars are similar for the other values of 
$\Omega_M$.  Note the significant difference between the deeper Durham/AAT
data and the LBQS and HBQS data.  This may indicate that the choice of
$g_k(z)$ is not correct.
Figure a) shows results for cosmological models with $\Omega_\Lambda =0$ and
$\Omega_k = 1 - \Omega_M$.
Figure b) shows results for cosmological models with $\Omega_k =0$ and
$\Omega_\Lambda = 1 - \Omega_M$.
}
\end{figure}

\subsubsection{Evolution of the form $g_{k'}(z) \propto (1 + z) ^ {k'}$}
Figures 4a and 4b give results for the evolution parameter $k'$
for the same cosmological models as above.
The allowed ranges obtained from 
the different samples are much closer, thus this form
of evolution is shown to be a closer approximation to the actual
evolution than the exponential form.  
For the Einstein - de Sitter model, the best value of
evolution parameter is $k' = 2.58$ with a one 
$\sigma$ range of $k' \in [2.14,2.91]$.
As with the previous case, the values of $k'$
are somewhat less than the previously found value of $k' \approx 3$
of Caditz and Petrosian (1990).

\begin{figure}[htbp]
\leavevmode
\centerline{
\psfig{file=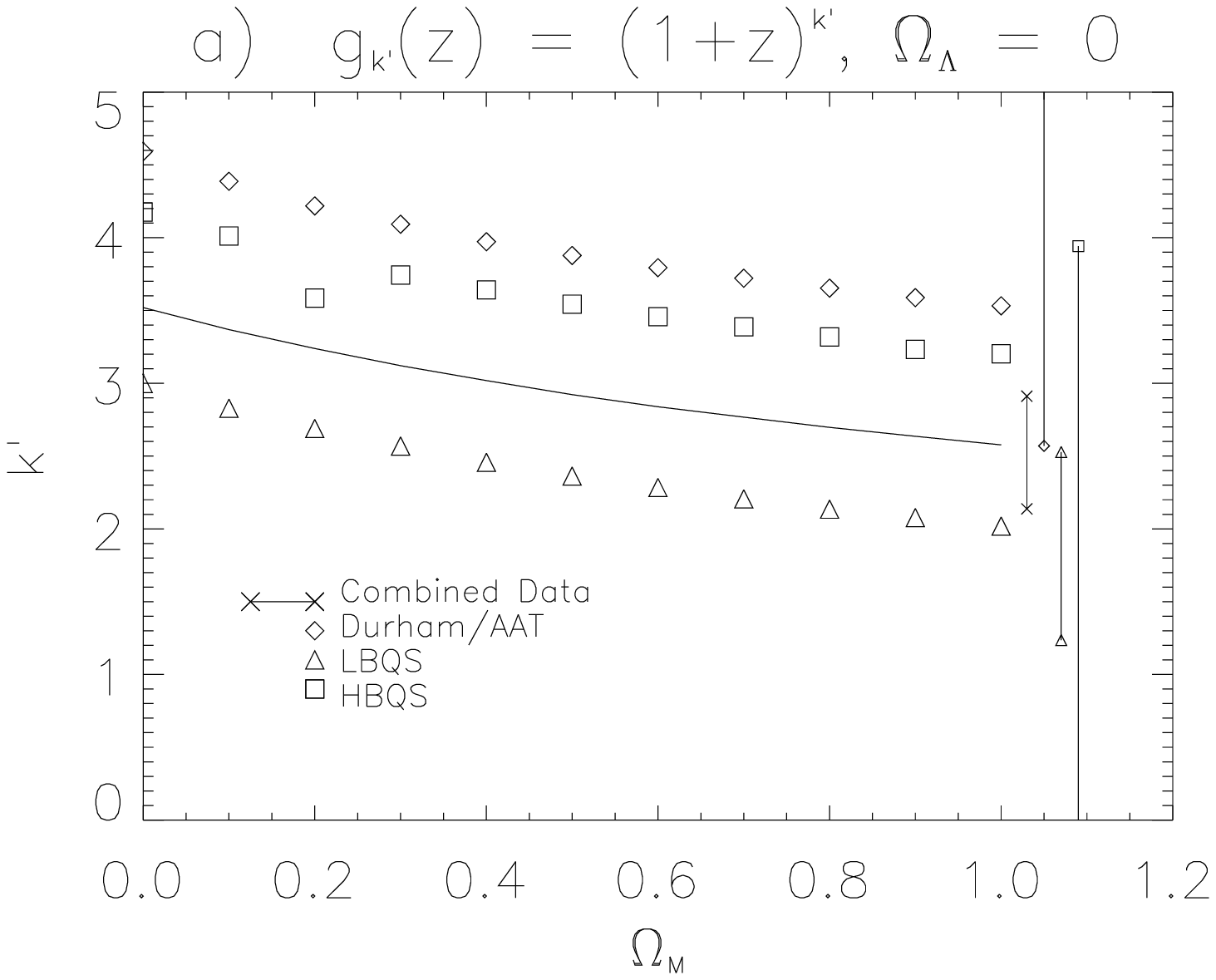,width=0.5\textwidth,height=0.4\textwidth}
\psfig{file=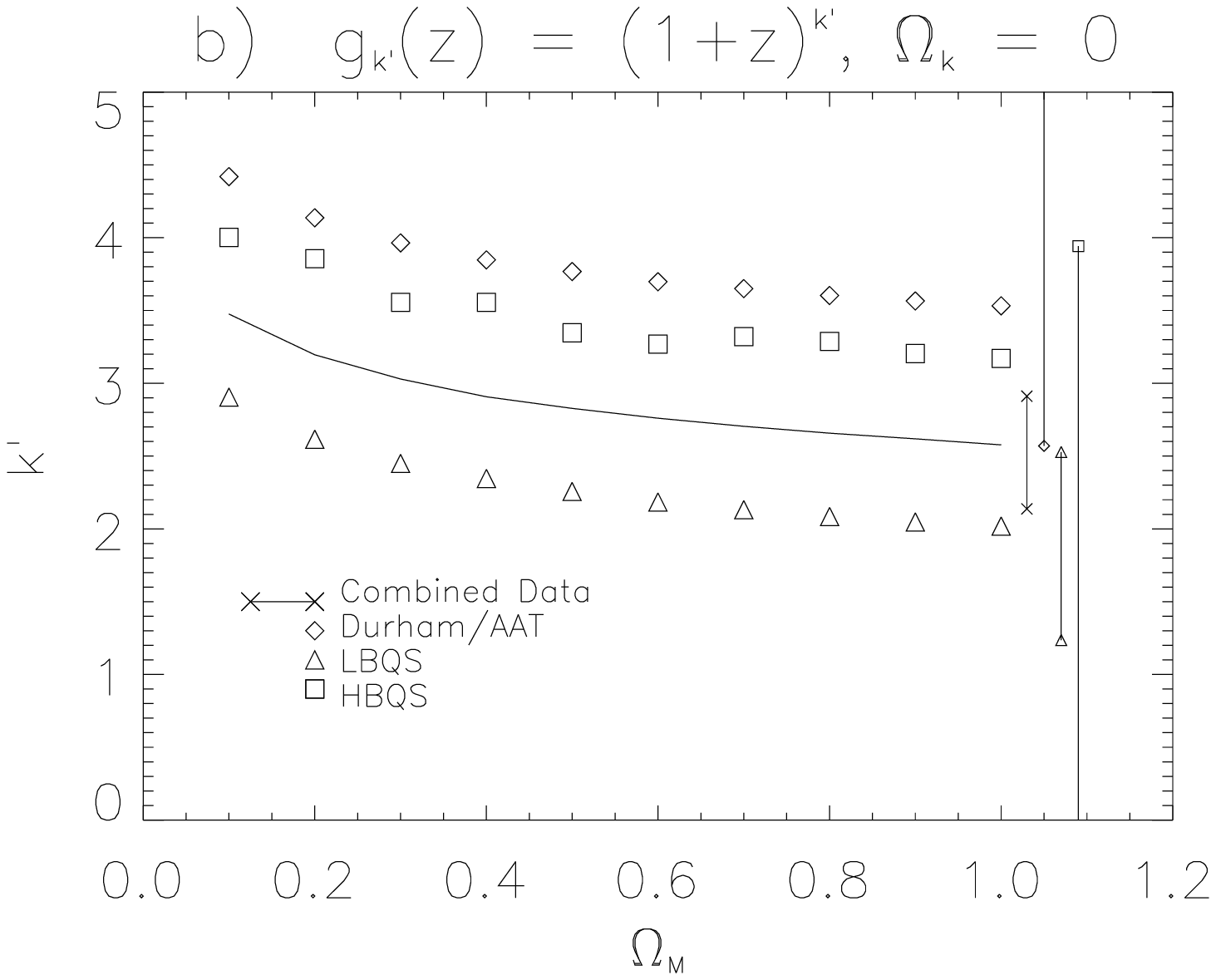,width=0.5\textwidth,height=0.4\textwidth}
}
\caption{
Same as Figure 3 with evolution 
of the form $g_{k'}(z) \propto (1 + z) ^ {k'}$.  The error bars nearly
overlap at 
the optimal value for the combined data, indicating that this is an 
better parameterization over the observed range of the $L - z$ 
plane.}
\end{figure}

It is clear that a better fit can be achieved with a different functional 
form for $g(z)$ with two or more parameters.
However, in what follows we assume the simpler form of evolution
$g(z) \propto (1+z)^{k'}$ with $k' (\Omega)$ the optimal value of $k'$
for a given cosmological model $\Omega$.
We therefore transform the data to $\{[L_{o \\ i},z_i]\}_{i=1}^N$ with 
$L_o(z,m,\Omega) = L(z,m,\Omega) / (1+z)^{k'}$ and apply the method of $\S 3$
to find non-parametric estimates for the density
functions $\rho(z)$ and $\psi(L_o)$.  This method now gives 
directly the cumulative functions 
$\sigma (z) = \int_0^z \rho (z) {dV \over dz} dz$ and
$\Phi (L_o) = \int _{L'}^\infty \psi(L') dL'$.

\subsection{The Density Evolution $\rho(z)$}
The cumulative density function $\sigma(z)$
is the total number of objects within the angular area of the survey
up to redshift $z$. 
If there is no density evolution, i.e. $\rho(z) = \rho_o$ is a constant,
then $\sigma \propto V$ where $V(z)$ is the co-moving volume 
up to redshift $z$. 
We determine $\sigma (z)$ and $\rho(z)$
using the new method for doubly truncated data. 
In order to determine if density evolution exists we fit $\sigma$ to $V$ by a 
simple power law $\sigma(z) \propto V^\lambda$, where
$\lambda \not = 1$ indicates the presence of density evolution.
If the density increases with redshift we expect $\lambda > 1$ and
if the density decreases with redshift we expect $\lambda < 1$.
Even if $\lambda = 1$, however, density evolution may be present:
the density may increase and decrease
in such a way as to cancel and give a fit of $\lambda = 1$.
Figures 5a and 5b show the variation of $\sigma$ and $\rho$
with $V$ for the combined
sample for three different cosmological models. 
The dotted lines show the best fits to the form 
$\sigma \propto [V(z,\Omega)]^\lambda$.
For the Einstein - de Sitter model (top curves in Figure 5)
we have $\lambda  = 1.19$,
indicating that the co-moving density increases with redshift roughly as 
$\sigma \propto V^{1.19}$.  
This would indicate a simple power law density evolution
$\rho \propto V^{0.19}$.  The density evolution shown in Figure 5b
exhibits this average behavior, but in detail is more complex:
the density increases more rapidly at low $z$,
reaches a plateau at $z\approx 2$ and
possibly decreases at higher $z$.
This behavior
will be discussed below in more detail and for higher redshift data.

\begin{figure}[htbp]
\leavevmode
\centerline{
\psfig{file=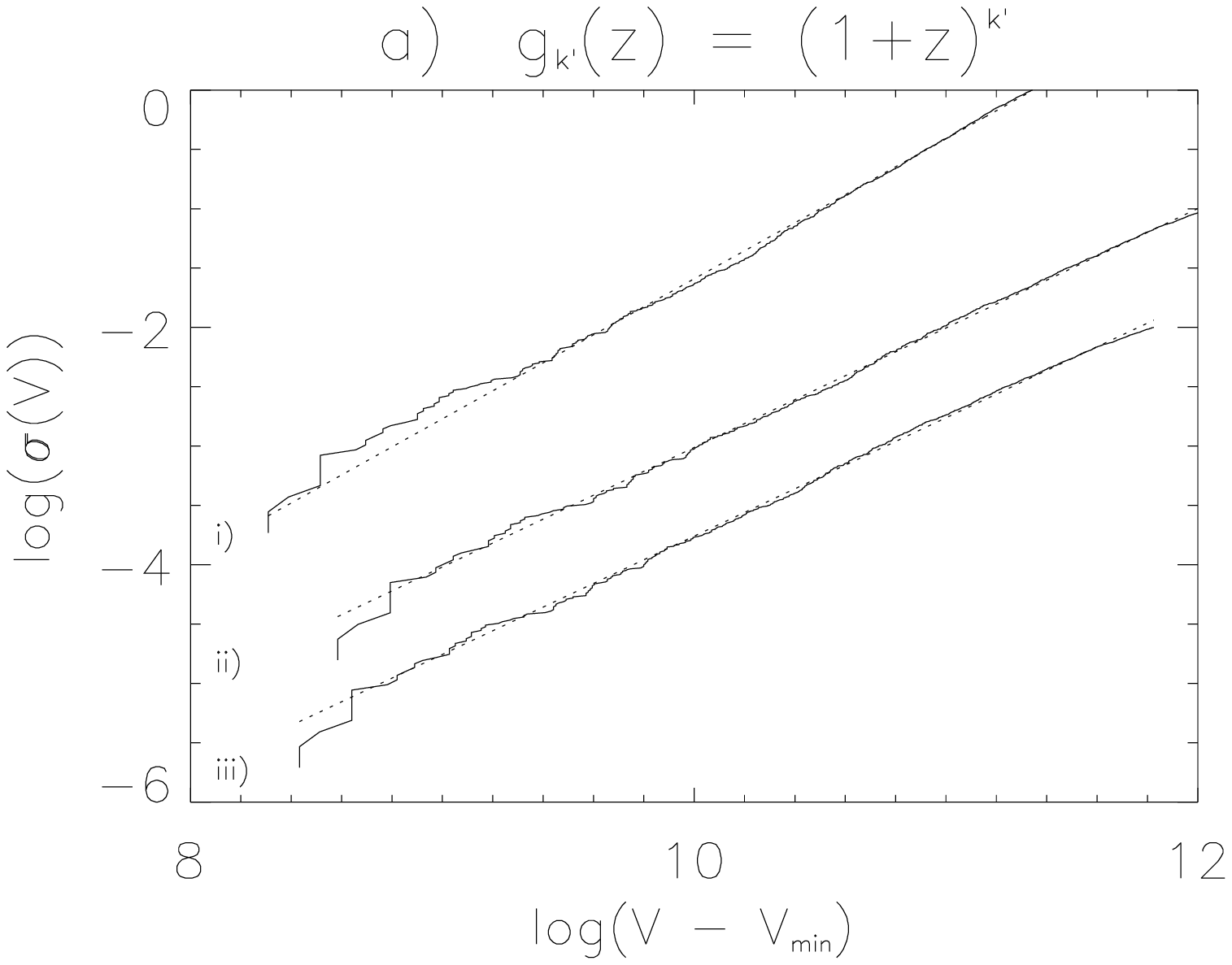,width=0.5\textwidth,height=0.4\textwidth}
\psfig{file=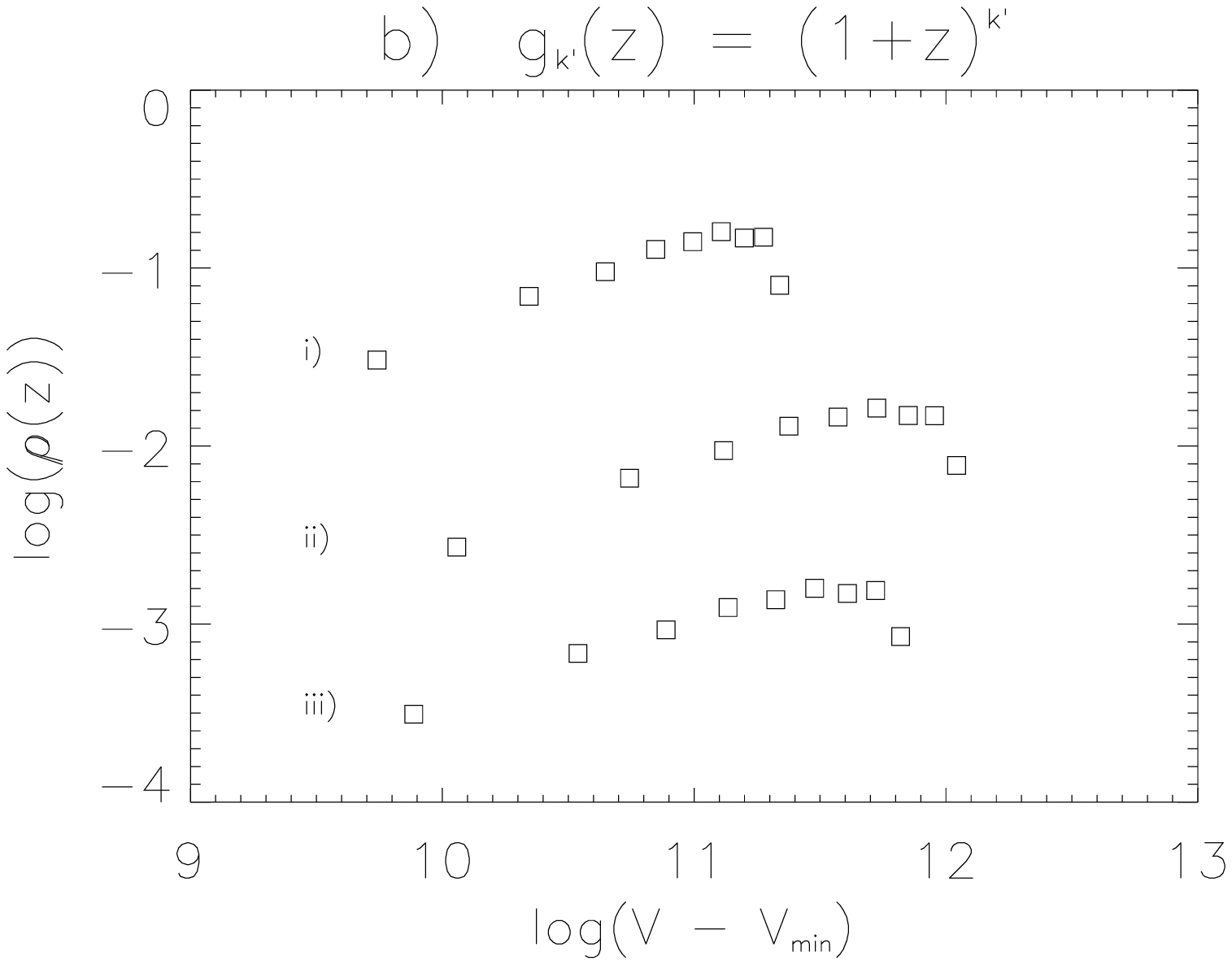,width=0.5\textwidth,height=0.4\textwidth}
}
\caption{
A plot of a) $\sigma = \int _0 ^z \rho (z') dz'$
and b) $\rho$
as a function of co-moving
volume $V(z,\Omega)$ for the combined data set for three different 
cosmological models.
Curves i), ii) and iii) show results for Einstein - de Sitter,
$\Omega_M = 0.15$, $\Omega_\Lambda = 0.85$ and 
$\Omega_M = 0$, $\Omega_\Lambda = 0$ models, respectively.
Vertical normalizations are arbitrary.
The straight dotted lines show the best fits to the 
parametric models $\sigma \propto V^\lambda$, with $\lambda=1.19$
for the Einstein - de Sitter model and $\lambda = 1$ for the other
two models.  
It is clear from b), however, that the model $\sigma \propto V^\lambda$
is only a rough approximation.  In fact, $\rho$ undergoes more complicated 
evolution, increasing and peaking before decreasing at higher redshift. 
The luminosity
evolution $g_{k'}(z) \propto (1 + z) ^ {k'}$, with the optimal value
of $k' = 2.58$, was used to remove the
correlation from the original data.
}
\end{figure}

As mentioned in \S 1, one cannot determine the evolution of sources 
(the function $\rho (z)$) and the evolution of the universe 
(the parameters $\Omega_i$)
simultaneously.  Given one of these (e.g. the cosmological model), 
the other (the density evolution) may be determined from the
data.  
The variation of $\lambda = d \ln \sigma / d \ln V$ with 
$\Omega_M$ for the two different classes of 
cosmological models is shown in Figures 6a and 6b.  
It is evident that there is a monotonic variation of $\lambda$ with
$\Omega_M$.  In particular, $\lambda = 1$
for the two cosmological models:
$\Omega_M \approx 0$, $\Omega_\Lambda \approx 1$
and $\Omega_M \approx 0.15$, $\Omega_\Lambda \approx 0.85$.  
This second set of parameters is quite close to those currently
favored by many observations.  As above, this does not imply the complete
absence of density evolution.  The lower two curves in Figure 5b
show the variation of $\rho$ for these two models.  Clearly, there
is less variation than for the Einstein - de Sitter model, but the general
behavior is similar.

\begin{figure}[htbp]
\leavevmode
\centerline{
\psfig{file=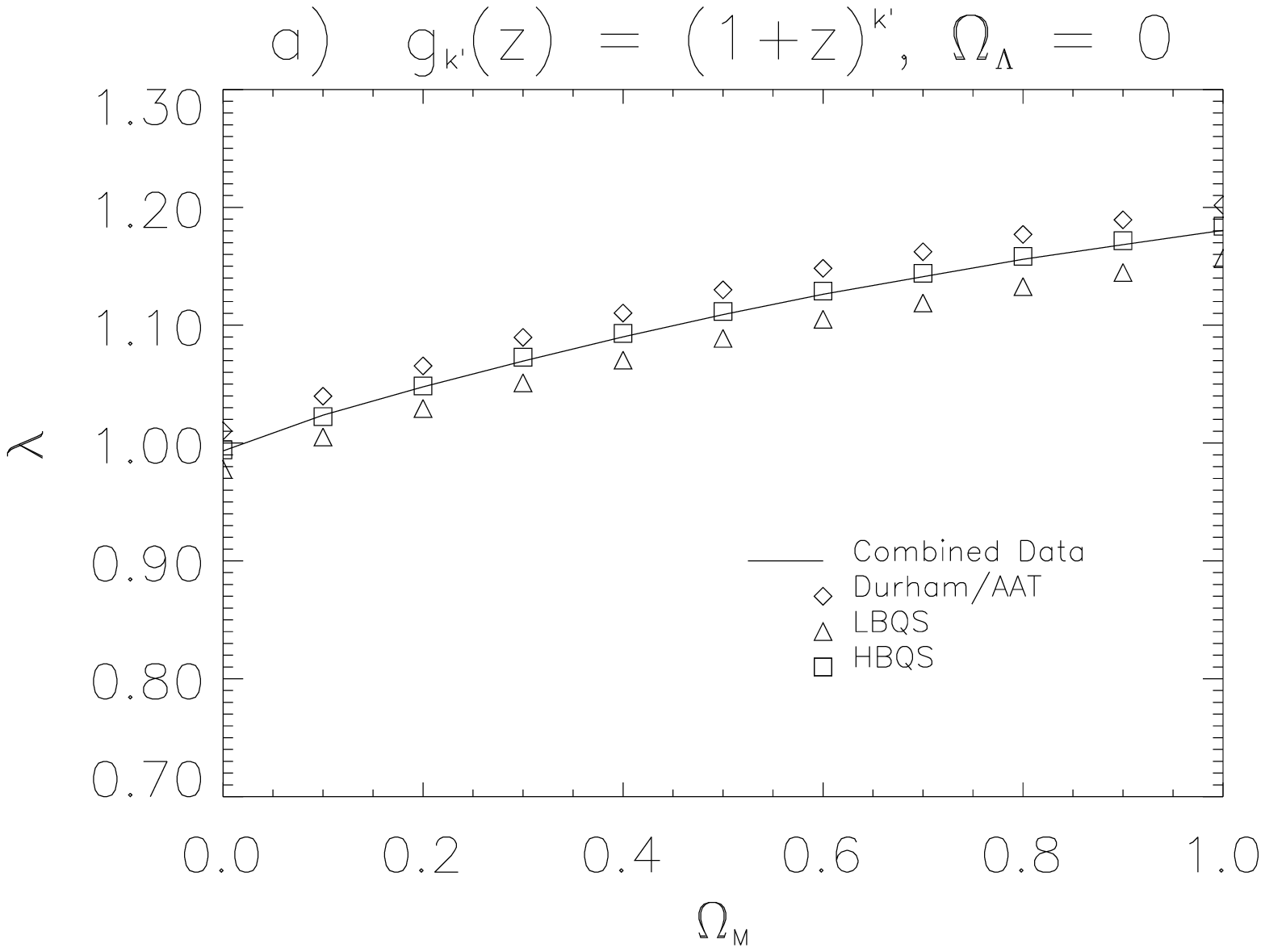,width=0.5\textwidth,height=0.4\textwidth}
\psfig{file=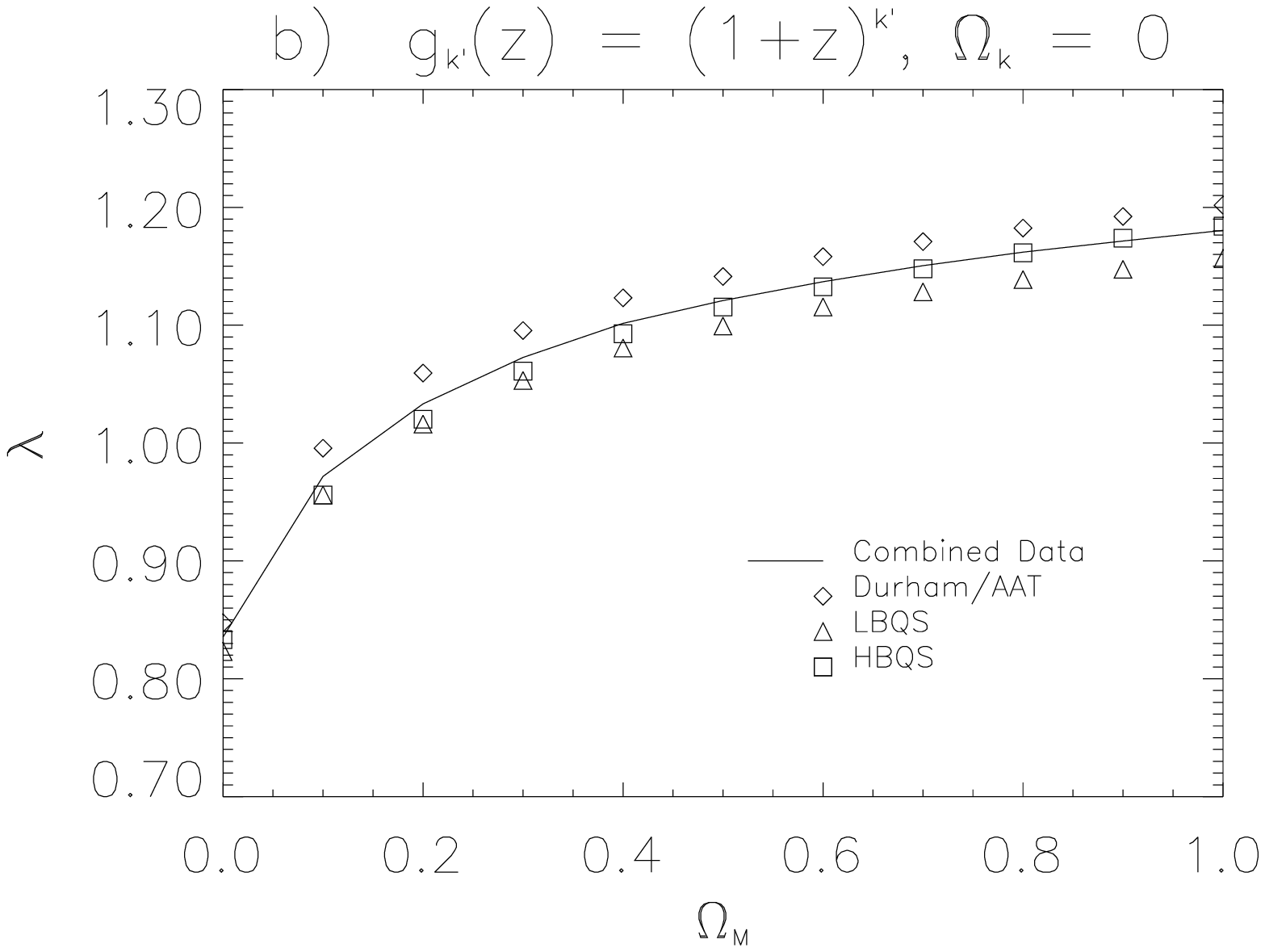,width=0.5\textwidth,height=0.4\textwidth}
}
\caption{The best fit for the power law index 
$\lambda = d \ln \sigma / d \ln V$ is shown as a function
of $\Omega_M$ for two different types of cosmological models. 
The index $\lambda$ is defined by the
equation $\sigma(z) \propto V^\lambda$,
where $V(z,\Omega)$ is the co-moving volume of space 
contained within the sphere of radius $R(z)$.  A value of
$\lambda < 1$ corresponds to a decrease in density with redshift and
$\lambda > 1$ corresponds to an increase in density with redshift.
Pure luminosity evolution, $\lambda = 1$,
($\sigma \propto V(z)$) occurs for models without cosmological constant at
$\Omega_M \approx \Omega_\Lambda \approx 0$ and for models without 
curvature at
$\Omega_M \approx 0.15, \Omega_\Lambda \approx 0.85$. 
Figure a) shows results for cosmological models with
$\Omega_\Lambda = 0$.  Figure b) shows results for models with $\Omega_k = 0$. 
}
\end{figure}

To further analyze the variation of $\rho(z)$, in
Figure 7 we show our non-parametric
determination of $\rho(z)$ for the Einstein - de Sitter cosmological model.  
Two sets of results are given.  
The first of these (depicted by squares) shows 
$\rho(z)$ in the region $0.3 < z < 2.2$ from
the combined data.  The second (triangles) shows $\rho(z)$
in the region $0.3 < z < 3.3$ from the
LBQS data (which do not have a high redshift cutoff at $z = 2.2$) alone.  
As evident in Figure 7, the density 
increases relatively slowly at low redshift 
($\rho \sim (1 + z)^{2.5}$) before reaching a peak at 
$z \approx 2$ and decreasing rapidly ($\rho \sim (1 + z)^{-5}$)
at higher redshift.
As discussed further in \S 5.5, 
the decrease in density present in this data at redshift of about 2 
is in agreement with high redshift survey results (Schmidt et al. 1995, 
Warren et al. 1994).

\begin{figure}[htbp]
\leavevmode\centering
\psfig{file=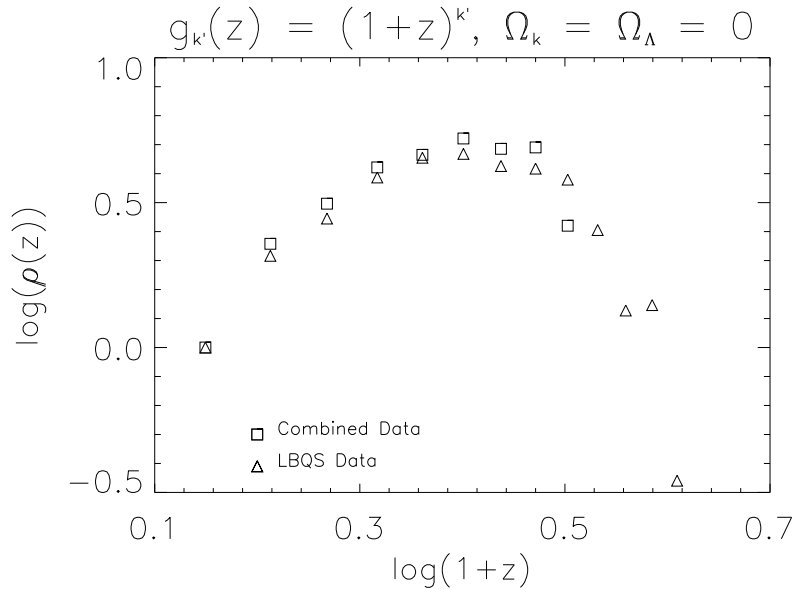,width=0.7\textwidth,height=0.5\textwidth}
\caption{
The non-parametric results for the density function $\rho(z)$
for the Einstein - de Sitter cosmological model.  
Clearly, a simple model $\rho \propto (1 + z)^\mu$ will not 
adequately describe the density evolution.  The squares show the results
for the combined data in the region $0.3 < z < 2.2$ and the
triangles show the results for the LBQS data in the region
$0.3 < z < 3.3$. The density peaks at $z \approx 2$.
}
\end{figure}

\subsection{The Luminosity Function $\psi(L_o)$}

In a similar fashion we may obtain the cumulative luminosity function 
$\Phi(L_o)$ from the uncorrelated data set $\{L_o,z\}$.  Figures 8a and 8b 
show $\Phi(L_o)$ for the combined data set with $k'=2.58$,
along with the best fits to a double power law form
\beq\label{eq:doublepow}
\Phi(L_o) = {\Phi_o \over (L_o / L_\ast)^{k_1} + (L_o / L_\ast)^{k_2}} .
\eeq
We used the cosmological
models $\Omega_M=1$, $\Omega_\Lambda =0$ (Einstein - de Sitter model)
and $\Omega_M = 0.15$, $\Omega_\Lambda = 0.85$
(pure luminosity evolution with cosmological constant).
In both cases, the results for the combined samples exhibit roughly
double power law dependence on $L_o$ with similar values of the fitting
parameters.
The primary differences between the two models are that the
Einstein - de Sitter model gives a slightly gentler slope
above the break luminosity, $k_2 = 3.17$ as opposed to $k_2 = 3.59$,
and a lower break luminosity,
$L_\ast = 6.19 \times 10^{29}$ erg / (sec Hz) as opposed to 
$L_\ast = 9.48 \times 10^{29}$ erg / (sec Hz).

\begin{figure}[htbp]
\leavevmode
\centerline{
\psfig{file=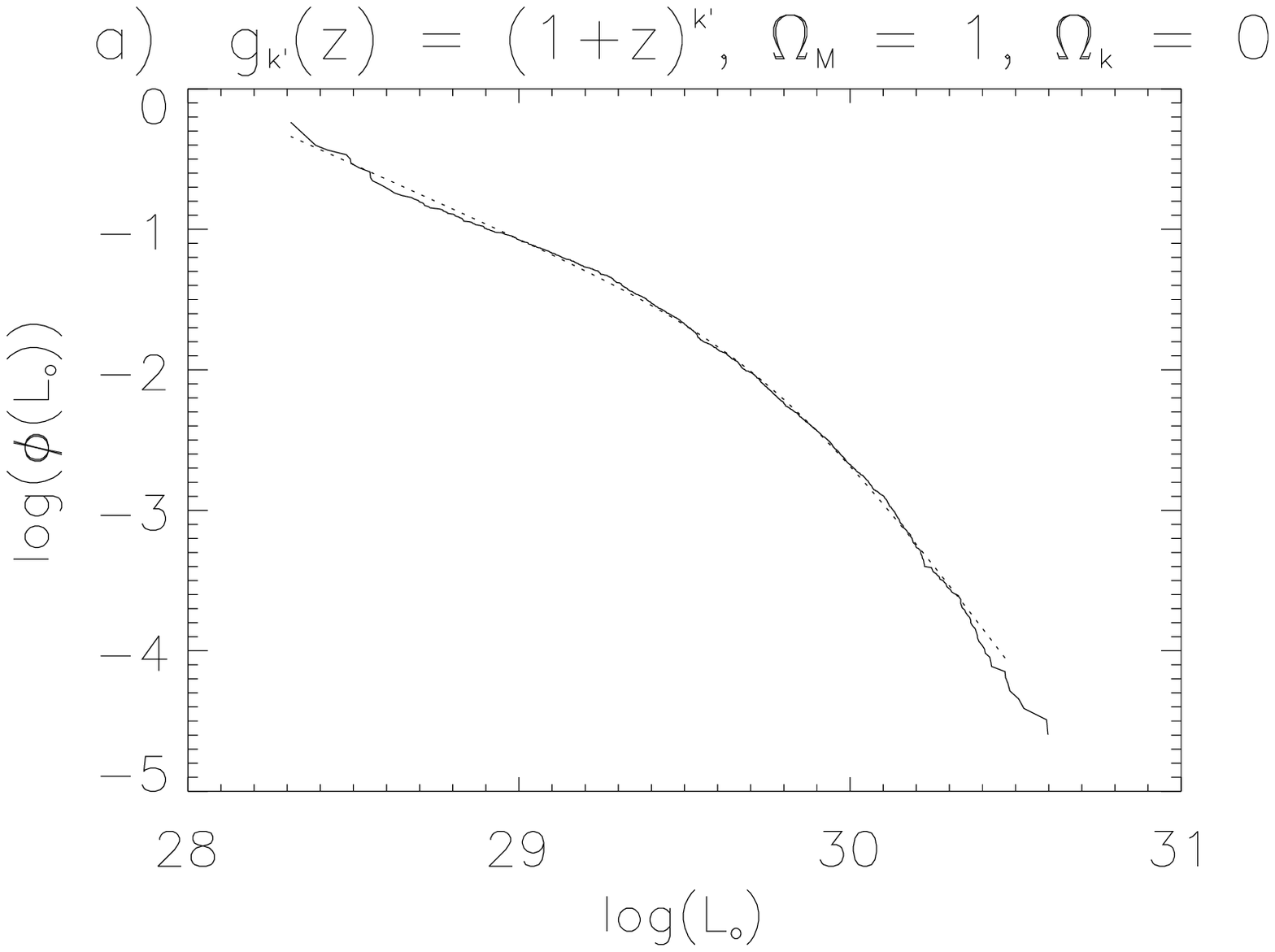,width=0.5\textwidth,height=0.4\textwidth}
\psfig{file=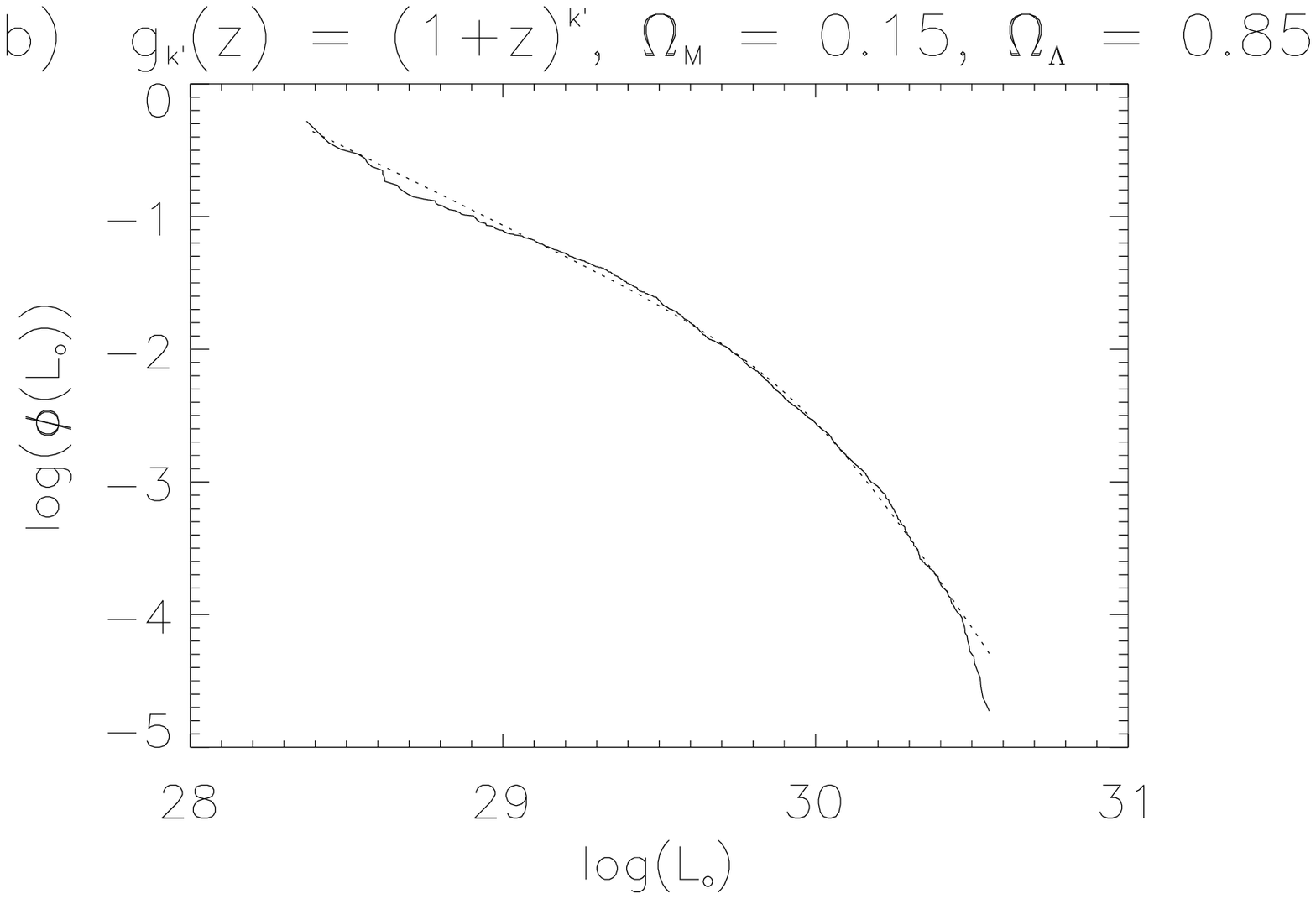,width=0.5\textwidth,height=0.4\textwidth}
}
\caption{The cumulative luminosity function 
$\Phi(L_o) = \int _{L_o} ^\infty \psi(L') dL'$
for the combined data set is shown as a function of $L_o = L / (1+z)^{k'}$ 
for two different cosmological models.
The dotted curve is the best fit to a double power law model,
equation (28).
Figure a) shows results for the Einstein - de Sitter model.
In this case, the best fit to the double
power law form has break luminosity 
$L_\ast = 6.19 \times 10^{29}$ erg / (sec Hz) and
power law indices $k_1 = 1.05$ and $k_2 = 3.17$.
Figure b) shows results for a cosmological model with 
$\Omega_M = 0.15$ and $\Omega_\Lambda = 0.85$.  The best fit for this 
model has break luminosity 
$L_\ast = 9.48 \times 10^{29}$ erg / (sec Hz) and
power law indices $k_1 = 1.16$ and $k_2 = 3.59$.
} 
\end{figure}

We check for possible variation in the shape of $\psi(L_o)$ by 
dividing the data into three redshift bins: 
$ 0.3 < z < 0.86 $, $ 0.86 < z < 1.48 $, and $ 1.48 < z < 2.2$.  
We then find the differential
luminosity function $\psi(L_o)$ for these three 
redshift bins, first assuming no luminosity evolution 
($g(z) = $ constant) and then assuming luminosity evolution
$g(z) \propto (1+z)^{k'}$.  
These luminosity functions (with arbitrary vertical normalization)
are shown in Figures 9a and 9b, respectively.
The presence of a strong shift to higher luminosities is clearly evident
for $g(z) = $ constant.  However, when the
evolution $g(z) \propto (1+z)^{k'}$ is taken out 
the luminosity function seems to
exhibit little variation; the slopes appear roughly the same at
low $L_o$ and high $L_o$ and the break luminosity does not vary as much
with redshift.
Although imprecise, these results indicate that our choice of
$g(z)$ removes most of the variation of the parameters $\alpha_i$
with redshift, i.e. the shape of the luminosity function is 
almost invariant.

\begin{figure}[htbp]
\leavevmode
\centerline{
\psfig{file=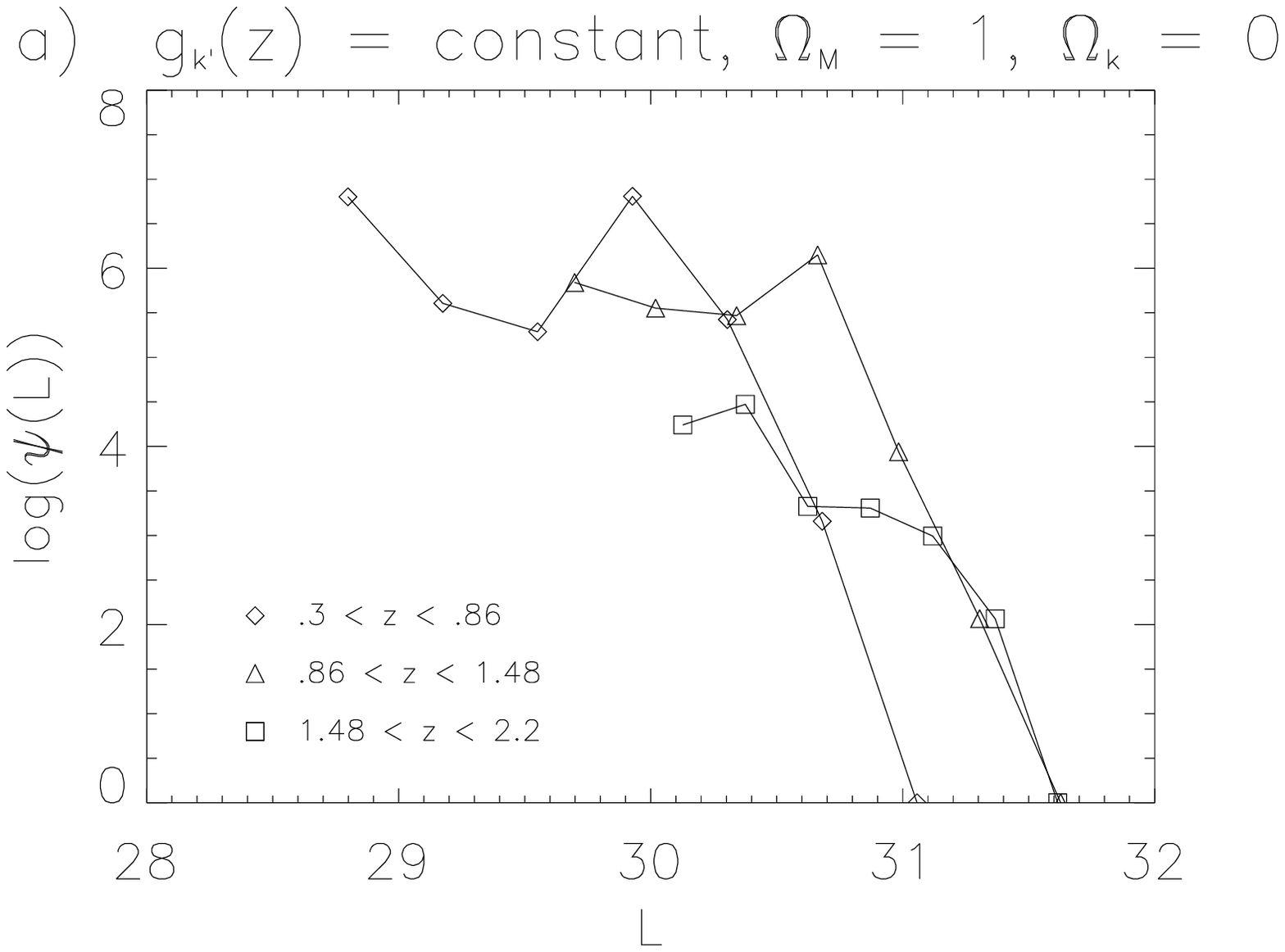,width=0.5\textwidth,height=0.4\textwidth}
\psfig{file=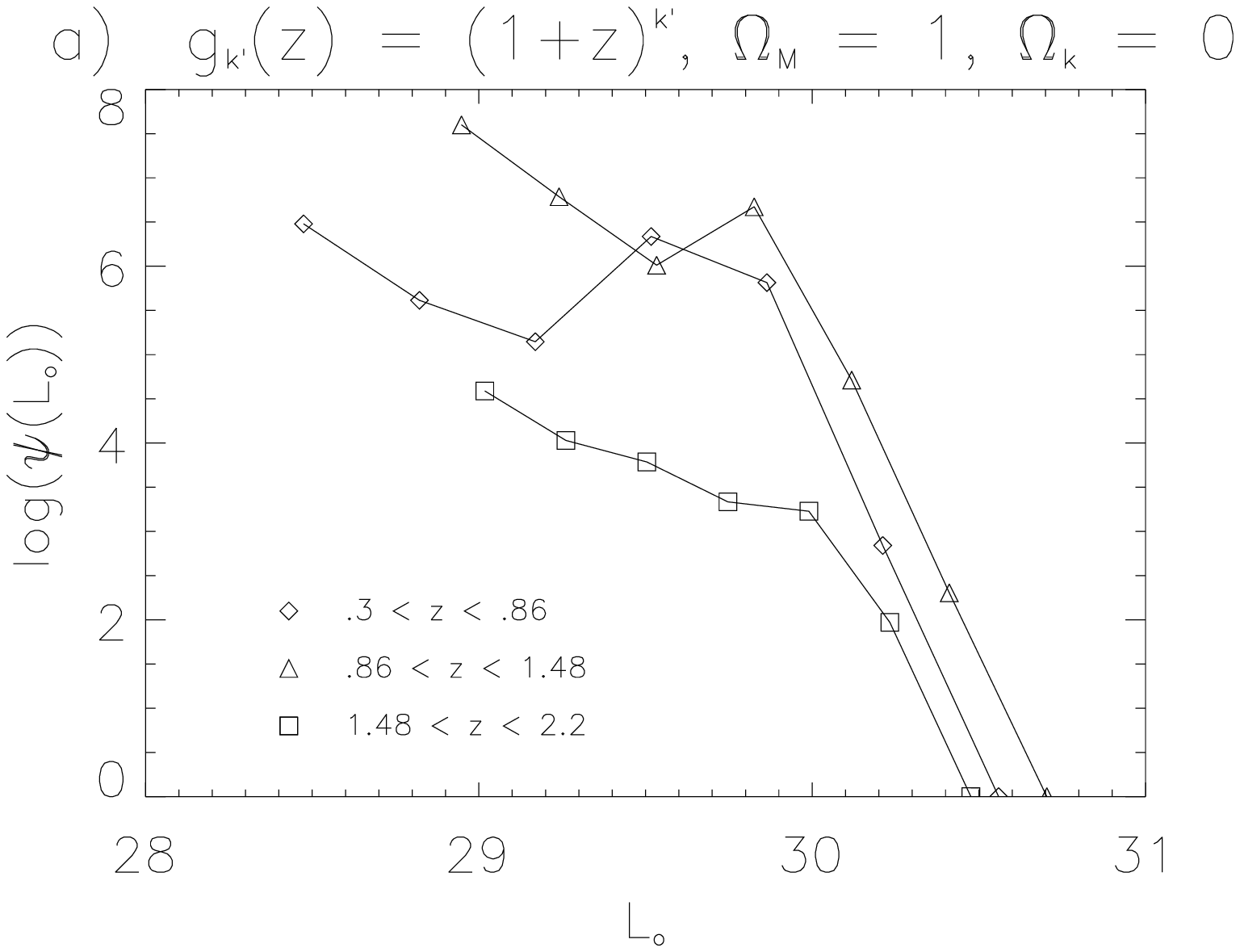,width=0.5\textwidth,height=0.4\textwidth}
}
\caption{
The luminosity function $\psi(L_o)$
for three different redshift bins (containing equal
numbers of sources) for the Einstein -
de Sitter cosmological model.  Vertical normalization is
arbitrary.
Figure a) shows results assuming no luminosity evolution, 
$g(z) = $ constant.  Figure b) shows results assuming 
luminosity evolution $g(z) \propto (1+z)^{k'}$ with the optimal value of
the parameter $k' = 2.58$ found above.
The second figure shows small variation in the shape of
$\psi(L_o)$ with redshift, indicating that most of the variation of
the parameters $\alpha_i$ is removed by the choice of luminosity evolution
$g(z) \propto (1+z)^{k'}$.
As expected, the high luminosity end can be fitted to a power law with index 
$k_2 + 1 \simeq 4$.
} 
\end{figure}

\subsection{The Luminosity Density ${\cal L} (z)$}

Finally, we determine the luminosity density function ${\cal L} (z)$.
This quantity is defined as the total rate of energy production by quasars
in the optical range as a function of redshift;
${\cal L} (z) = \int_0^\infty L \psi(L) dL$.
If the shape of the luminosity function is invariant then
${\cal L} (z) \propto \rho(z) g(z)$.
This rate depends in a complicated way on the distribution of masses
of the central black holes and the variation of the accretion rate,
both of which are related to the formation of galaxies and their evolution
through mergers or collisions.
Using the above results we can evaluate ${\cal L}(z)$ up to a redshift of
$2.2$.  We extend this further to a redshift of $3.3$ using the LBQS
data, which is claimed to be complete up to this redshift.  
These results, with arbitrary 
vertical normalization, are shown in Figure 10.
We first note the good agreement in the $z < 2$ region, indicating that
perhaps the LBQS result at higher redshift is a representative behavior.
We may also use high redshift surveys of quasars 
(Schmidt et al. 1995, Warren et al. 1994) to study ${\cal L}(z)$ in 
this range.  Unfortunately, the selection of high $z$ quasars in these
samples is more complicated and the subsequent analyses involve more 
assumptions.  For example, Schmidt et al. use the $V / V_{max}$ method
to determine $\rho(z)$, tacitly assuming that $g(z) = $ constant (as
well as $\alpha_i = $ constant) so that ${\cal L} (z) \propto \rho (z)$.
We show these results (again with arbitrary vertical normalization)
in Figure 10.  These results agree with the general trend of decline in 
${\cal L} (z)$ at high redshifts.

\begin{figure}[htbp]
\leavevmode\centering
\psfig{file=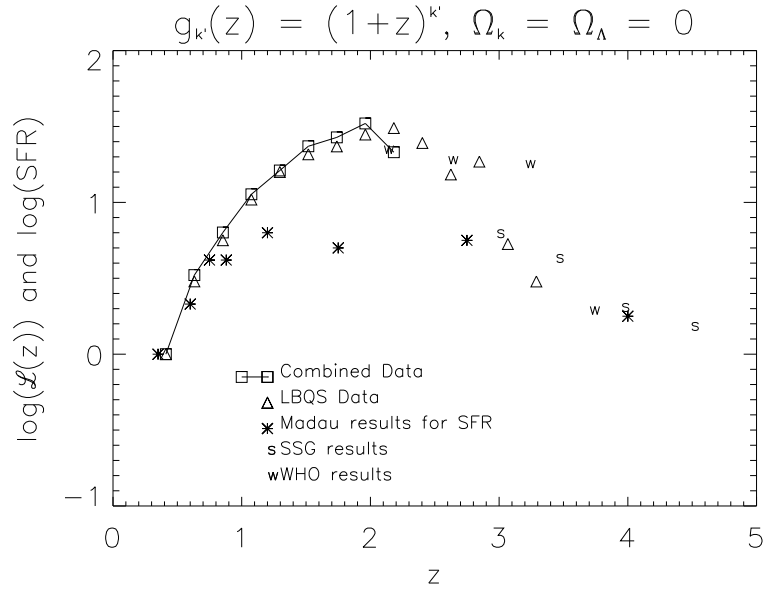,width=0.7\textwidth,height=0.5\textwidth}
\caption{The luminosity density ${\cal L} (z)$ is shown as a function of
redshift for the Einstein - de Sitter cosmological model. 
The squares and triangles give our results for the combined data and the 
LBQS data, respectively.
The high redshift results of Schmidt et al. (1995) and 
Warren et al. (1994) are given by
the letters ``s'' and ``w'', respectively.
The vertical normalizations are arbitrary.
All of the above results indicate that ${\cal L}$ peaks
somewhere in the region $z \approx 2$.
The stellar formation rate (SFR) as a function of redshift, 
as found by Madau (1997), is given by the asterisks.  Although
the SFR also exhibits the same general behavior as ${\cal L} (z)$, the
peak occurs at a lower redshift and appears somewhat broader.
}
\end{figure}

It has been claimed (Cavaliere and Vittorini 1998,
Shaver et al. 1998) 
that this rise and fall of ${\cal L} (z)$ 
with redshift is similar to the behavior of the star formation rate
(SFR), which has recently been extended to high redshift 
(see, e.g. Madau 1997).
We have shown this rate in Figure 10 as well.  Although the general trend
of rise and fall of the SFR and ${\cal L} (z)$ is the same,
there is considerable difference in the detailed variation.
The similarity may indicate some relation between the SFR and the 
feeding of the central engine of the quasars (e.g. both are affected by
mergers).  However, considering the many differences between star
formation and the generation of energy by quasars, the observed difference
between the SFR and ${\cal L} (z)$ in Figure 10 is not surprising.

\section{SUMMARY AND CONCLUSIONS}\label{sec:discussion}

Although there have been several analyses of quasar evolution in the past,
our results differ from these previous results in two important respects:

$\bullet$ We have used non-parametric statistical methods 
for multiply truncated 
data that allow us to combine samples with different selection criteria.

$\bullet$ We have used the data to study models of the luminosity function
that take into account 
both luminosity evolution $g(z)$ and density evolution $\rho (z)$. 

The new non-parametric statistical methods differ from those used in the past 
in the following ways:

$\bullet$ No binning is required and most of the characteristics of the
distribution functions are determined non-parametrically.

$\bullet$ The methods are not limited to simply truncated data such
as that found in flux limited surveys and can account for selection
biases in generally truncated data where each data point has different 
truncation limits.  
In particular, these methods can treat samples with both upper 
and lower flux limits and redshift limits.

$\bullet$ This versatility allows one to combine data from different surveys 
with different selection criteria.

$\bullet$ The first of our techniques, a 
generalized non-parametric test of independence, allows one to 
determine the degree of correlation between luminosity and redshift,
giving an indication of the luminosity evolution in the luminosity 
function.  The evolution may then be determined
parametrically.

$\bullet$ The second of our techniques provides a non-parametric estimate for 
the univariate distributions in redshift and luminosity, 
i.e. the co-moving density evolution and the local luminosity function,
respectively. 

We have applied these methods to the combined data from several
large surveys and determined the luminosity evolution $g(z)$,
the density evolution $\rho(z)$ and the luminosity function 
$\psi(L_o = L/g(z))$ of the generalized luminosity function of equation
(\ref{eq:gle}) for flat ($\Omega_k = 0$ and $\Omega_M = 1 - \Omega_\Lambda$)
and zero cosmological constant ($\Omega_\Lambda = 0$ and 
$\Omega_M = 1 - \Omega_k$) cosmological models.
We assume a shape invariant luminosity function, $\alpha_i = $ constant.
More complex luminosity functions, $\alpha_i \not = $ constant,
or those with luminosity dependent density evolution, etc., can be tested if 
the simpler prescription used here is not consistent with all of the data.
We found that the scenario of equation (\ref{eq:gle}) provides an 
adequate description of the existing data.

Our results may be summarized as follows:

$\bullet$ We found a strong correlation between luminosity and redshift,
indicating the presence of rapid luminosity evolution.

$\bullet$
The parametric model of luminosity evolution $(1+z)^{k'}$ provides a better
description of the data than the model $e^{k t(z)}$, although neither 
parameterization 
perfectly removes the correlation in all areas of the $L - z$ plane.
In order to better model this evolution future analyses of quasar evolution
could consider other parametric forms,
including those with more than one free parameter. 

$\bullet$  The cumulative co-moving density of quasars may be 
approximately modeled as $\sigma (z) \propto V^\lambda$, 
where the value of $\lambda$ depends on the cosmological model.  For example, 
$\lambda = 1.19$ for the Einstein - de Sitter model
and $\lambda = 1$ for the cosmological models with 
$\Omega_M \approx 0$, $\Omega_k \approx 1$ and
$\Omega_M \approx 0.15$, $\Omega_\Lambda \approx 0.85$.
This simple parameterization $\sigma \propto V^\lambda$ 
does not describe precisely the variation of $\rho$ with 
redshift.
When examined in greater detail, the co-moving density shows a relatively
slow increase ($\rho \sim (1 + z)^{2.5}$) for low redshifts and a
rapid decline ($\rho \sim (1 + z)^{-5}$) for $z > 2$
(for the Einstein - de Sitter model).  This is in
qualitative agreement with the observed density evolution of high
$z$ quasars (Schmidt et al. 1995, Warren et al. 1994).  
Qualitatively similar behavior is found even for models that show no
overall density evolution ($\lambda = 1$).
A more rigorous comparison cannot be made at this stage because the analysis
of the high $z$ data ignores possible $L - z$ correlation and assumes pure
density evolution.  

$\bullet$
The cumulative local luminosity function 
$\Phi(L_o) = \int _{L_o} ^ \infty \psi(x) dx$
has a double power law form.
In the Einstein - de Sitter model the break luminosity is
$L_\ast = 6 \times 10^{29}$ erg / (sec Hz) and the low and high luminosity
power law indices are 
$k_1 = 1.05$ and $k_2 = 3.17$.
There appears to be little variation with redshift of the shape
of the cumulative and differential luminosity functions, thus
the $\alpha_i$ = constant prescription seems adequate.  
With more data one could determine precisely
the variation with redshift of the shape of 
$\phi(L_o)$.

$\bullet$ The above description of the luminosity function allows us to 
determine the rate of energy generation per unit co-moving volume of
quasars as a function of redshift.  We show that this function 
${\cal L}(z) \propto \rho (z) g(z)$ increases rapidly with $z$ at low
redshift, peaks around $z \approx 2$ and then decreases.
This is also in rough agreement with the high $z$ survey results mentioned
above.
This variation of ${\cal L}(z)$ is similar to
but significantly different from recent determinations of the star
formation rate.

\acknowledgements
We would like to thank Bradley Efron for his invaluable help with the
statistical methods used in this paper, Paul Hewett for help with the analysis
of the LBQS data and Nicole Lloyd for helpful discussions.  A. M. would like
to acknowledge support from the Stanford University Undergraduate 
Research Opportunity program.

\end{document}